\documentclass[twocolumn,showpacs,preprintnumbers,amsmath,amssymb,prb,superscriptaddress]{revtex4-1}
\usepackage{multirow}
\usepackage{amsfonts}
\usepackage[english]{babel}
\usepackage[T1]{fontenc}
\usepackage{times}
\usepackage{mathrsfs}
\usepackage{graphicx}
\usepackage{dcolumn}
\usepackage{bm}
\usepackage[colorlinks,bookmarks=true,citecolor=blue,linkcolor=red,urlcolor=blue]{hyperref}
\usepackage[tight, FIGTOPCAP, hang, raggedright, nooneline]{subfigure}

\begin{document}

\title{Phase diagram of $\nu=\frac{1}{2}+\frac{1}{2}$ bilayer bosons with inter-layer couplings}

\author{Zhao Liu}
\affiliation{Dahlem Center for Complex Quantum Systems and Institut f\"ur Theoretische Physik, Freie Universit\"at Berlin, Arnimallee 14, 14195 Berlin, Germany}
\affiliation{Department of Electrical Engineering, Princeton University, Princeton, New Jersey 08544, USA}
\author{Abolhassan Vaezi}
\affiliation{Department of Physics, Stanford University, Stanford, California 94305, USA}
\affiliation{Department of Physics, Cornell University, Ithaca, New York 14853, USA}
\author{C\'ecile Repellin}
\affiliation{Max-Planck-Institut f\"ur Physik komplexer Systeme, 01187 Dresden, Germany}
\author{Nicolas Regnault}
\affiliation{Department of Physics, Princeton University, Princeton, New Jersey 08544, USA}
\affiliation{Laboratoire Pierre Aigrain, Ecole Normale Sup\'erieure-PSL Research University, CNRS, Universit\'e Pierre et Marie Curie-Sorbonne Universit\'es, Universit\'e Paris Diderot-Sorbonne Paris Cit\'e, 24 rue Lhomond, 75231 Paris Cedex 05, France}

\date{\today}

\begin{abstract}
We present the quantitative phase diagram of the bilayer bosonic fractional quantum Hall system on the torus geometry at total filling factor $\nu=1$ in the lowest Landau level. We consider short-range interactions within and between the two layers, as well as the inter-layer tunneling. In the fully polarized regime, we provide an updated detailed numerical analysis to establish the presence of the Moore-Read phase of both even and odd numbers of particles. In the actual bilayer situation, we find that both inter-layer interactions and tunneling can provide the physical mechanism necessary for the low-energy physics to be driven by the fully polarized regime, thus leading to the emergence of the Moore-Read phase. Inter-layer interactions favor a ferromagnetic phase when the system is $SU(2)$ symmetric, while the inter-layer tunneling acts as a Zeeman field polarizing the system. Besides the Moore-Read phase, the $(220)$ Halperin state and the coupled Moore-Read state are also realized in this model. We study their stability against each other.

\end{abstract}

\maketitle

\section{Introduction}
Elegant approaches to create topologically ordered quantum states have been proposed starting from a given parent one. These techniques could be useful to engineer a richer topological order or to potentially inherit from the parent state some non-universal properties such as a large gap. An example that has recently drawn much attention is the anyon condensation\cite{Bais-PhysRevB.79.045316,Barkeshli-PhysRevLett.105.216804}. In this context, condensing some of the bosonic excitations of a given topological phase leads to a simpler (or equally rich) topological order. Conversely, the projective construction~\cite{Cappelli-1999CMaPh.205..657C, Froehlich-2000cond.mat..2330F,cappelli-2001NuPhB.599..499C,Barkeshli-PhysRevB.81.155302,Repellin-PhysRevB.92.115128} starts from multiple copies of a simple topological phase hosting for example only Abelian excitations, to generate a new one that could potentially have a more complex topological order, involving non-Abelian excitations.

The projective construction can be thought of as several layers of a topological state that we symmetrize (or anti-symmetrize) over the layer degree of freedom. When the topological order can be described by conformal theories [such as for several fractional quantum Hall (FQH) model wave functions], this construction is related to the so-called coset/orbifold projections~\cite{Cappelli-1999CMaPh.205..657C, Froehlich-2000cond.mat..2330F,cappelli-2001NuPhB.599..499C,Barkeshli-PhysRevB.81.155302}. One simple example of the coset projection is based on two copies of the bosonic Laughlin $\nu=\frac {1}{2}$ state\cite{Laughlin-PhysRevLett.50.1395} leading to the Moore-Read (MR) state\cite{Moore1991362} once symmetrized \cite{Fradkin-1998NuPhB,Fradkin-1999NuPhB}. Another similar example, known as orbifolding, is based on two copies of the fermionic $\nu=\frac{1}{3}$ Laughlin state where anti-symmetrization yields the $\mathbb{Z}_4$ Read-Rezayi state. While being mathematically well defined, the symmetrization (or anti-symmetrization for fermionic systems) is not a physical process. Still, it was argued in Ref.~\onlinecite{read-green-PhysRevB.61.10267} that tunneling between layers might play the same role.

In this article, we discuss the phase diagram of a bilayer bosonic FQH system in the presence of inter-layer interactions and tunneling at total filling factor $\nu=\frac{1}{2}+\frac{1}{2}$. For that purpose, we use exact diagonalizations on the torus geometry. In each layer the particles interact via a contact interaction. When the two layers are decoupled, each of them is at filling factor $\nu=\frac{1}{2}$ leading to two copies of the $\nu=\frac{1}{2}$ Laughlin state. Applying the projective construction would allow to recover the MR state\cite{Repellin-PhysRevB.92.115128}. If all the bosons were located in the same layer thus having effectively a single layer at filling factor $\nu=1$, strong numerical evidence\cite{Cooper-PhysRevLett.87.120405,regnault-PhysRevLett.91.030402,Chang-PhysRevA.72.013611,PhysRevB.69.235309,PhysRevB.76.235324} has shown that the emerging phase would also be described by the MR state.

Recent works have considered such a setup either on different geometry and a slightly long-range interaction\cite{Wu-PhysRevB.87.245123,Moller-PhysRevB.90.235101,PhysRevA.91.063623} or using a lattice analogue via two copies of fractional Chern insulators\cite{PhysRevB.90.245401,Zhu-PhysRevB.91.245126}. Other studies have also considered such a system at larger filling factors in the context of the non-Abelian spin-singlet state\cite{Ardonne-PhysRevLett.82.5096} or the integer quantum Hall effect for bosons\cite{Furukawa-PhysRevLett.111.090401,Wu-PhysRevB.87.245123,Regnault-PhysRevB.88.161106}. A similar setup was also considered for fermions to look for the emergence\cite{Papic-PhysRevB.82.075302} of the fermionic MR state starting from the $(331)$ Halperin state\cite{halperin1983theory} or more recently to study the possible realization of the $\mathbb{Z}_4$ Read-Rezayi\cite{Rezayi-2010arXiv1007.2022R,Barkeshli-PhysRevLett.105.216804,Peterson-PhysRevB.92.035103} state out of two copies of the Laughlin $\nu=\frac{1}{3}$ state. These two cases are also instructive : neither features their respective non-Abelian state under full polarization (i.e. when all the particles are in the same layer) for a short range interaction (or even the Coulomb interaction) projected onto the lowest Landau level. Moreover, no signature was found for these respective non-Abelian states in the bilayer setup. However, this does not exclude the possibility to realize other non-Abelian states such as the Fibonacci\cite{Vaezi-PhysRevLett.113.236804,Liu-PhysRevB.92.081102} or interlayer Pfaffian\cite{Ardonne-2002,Geraedts-PhysRevB.91.205139} states.

The situation for the bosonic bilayer at $\nu=1$ is different. At large tunneling between the two layers, the system is effectively a single-component state\cite{Papic-PhysRevB.82.075302} with an effective interaction equal to the average of the intra-layer and the inter-layer interactions. Thus the physics of a single-layer bosonic FQH system at $\nu=1$ guarantees that the bilayer system hosts a MR phase at large tunneling. Similar arguments can be made with respect to the role of the inter-layer interaction. Indeed, choosing the same strength for the contact interaction within each layer and between layers, the bilayer recovers a full $SU(2)$ symmetry with respect to the layer index. If the low-energy physics is ``ferromagnetic'', it will be driven once again by the single-layer physics at $\nu=1$. Therefore, the aim of this article is not to check if the MR state might emerge in the bosonic bilayer FQH system but rather to give a {\it quantitative} phase diagram of this system, and look at the stability of the MR state. Other topological phases could also appear in such a bilayer system, such as the coupled MR (cMR) state proposed in  Ref.~\onlinecite{Hormozi-PhysRevLett.108.256809}. This state is akin to two chiral $p$-wave superconductors of composite fermions with a tunneling of Cooper pairs. While the cMR is the exact ground state of a combination of three-body intra-layer and two-body inter-layer interactions, we will show that it can accurately describe a region of the bilayer phase diagram.

The structure of this article is as follows. In Sec.~\ref{sec:bilayer}, we describe the bosonic bilayer model and briefly present the coupled Moore-Read state. We provide in Sec.~\ref{sec:MR} a full numerical analysis of the model when the system is completely polarized, i.e. the emergence of the MR state at filling factor $\nu=1$ for the bosonic fractional Hall effect with a two-body hardcore interaction in the lowest Landau level. Sec.~\ref{sec:phasediagraminter} describes the phase diagram when considering the two shortest-range pseudo-potentials for the inter-layer interaction. In particular we discuss the emergence and the stability of three distinct phases: the $(220)$ Halperin state, the MR state and the cMR state. Finally we consider the effect of a tunneling term between the two layers in Sec.~\ref{sec:phasediagramtunneling}.

\begin{figure*}
\centerline{\includegraphics[width=0.8\linewidth]{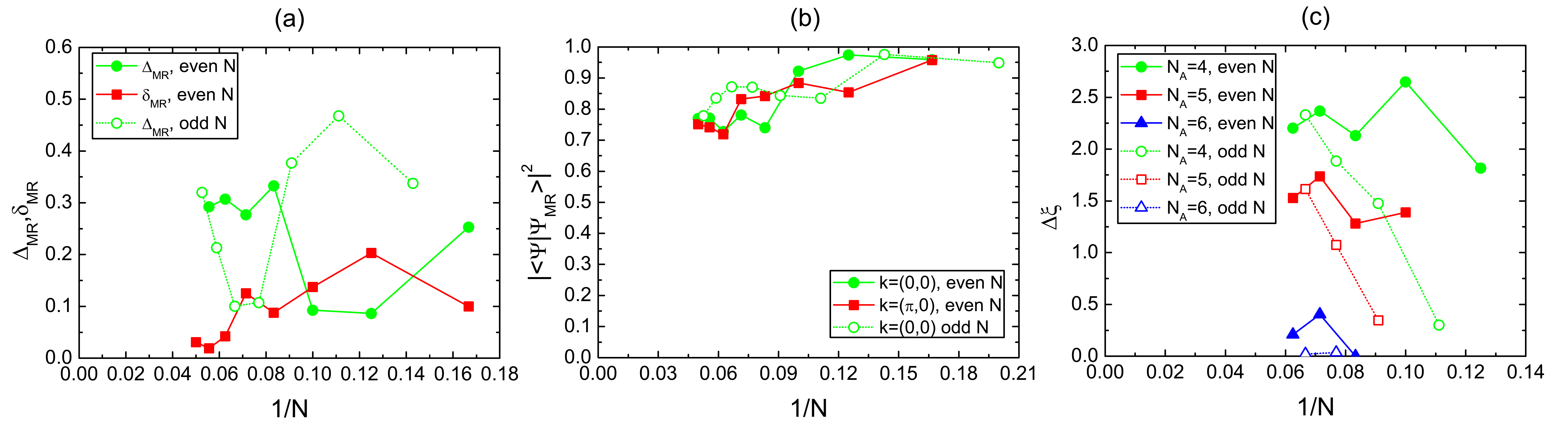}}
\caption{(Color online) (a) The finite-size scaling of the energy gap $\Delta_{\rm MR}$ for even $N$ (solid line) and odd $N$ (dotted line), and the ground-state splitting $\delta_{\rm MR}$ for even $N$ (solid line). The data from $N=6$ to $N=20$ are included (except the gap for $N=20$). (b) The ground-state overlap with exact MR states for even $N$ (solid line) and odd $N$ (dotted line) from $N=5$ to $N=20$. The overlap in the ${\bf k}=(0,\pi)$ sector is the same as that in the ${\bf k}=(\pi,0)$ sector due to the $C_4$ symmetry of the square torus. For $N=19$, the exact MR state cannot be built, thus we have used the very accurate approximation on the torus involving one Laughlin quasihole and one Laughlin quasiparticle\cite{Repellin-PhysRevB.92.115128}. (c) The finite-size scaling of the entanglement gap $\Delta_{\xi}$ in the $N_A=4$, $N_A=5$ and $N_A=6$ sectors for even $N$ (solid line) and odd $N$ (dotted line). The data are included only from $N=8$ to $N=16$ due to numerical limitations in the full diagonalization of reduced density matrices.}
\label{polarized}
\end{figure*}

\section{The $\nu=\frac{1}{2}+\frac{1}{2}$ Bosonic Bilayer}\label{sec:bilayer}

In this section, we first describe our model of bilayer bosons. The MR state, the $(2 2 0)$ Halperin state, and the coupled MR (cMR) state~\cite{Hormozi-PhysRevLett.108.256809} are three possible ground state candidates. Then we review the properties of the Halperin and cMR states, but leave the details of the MR state for the following sections.

\subsection{Model}\label{sec:model}

We consider a bilayer FQH system at total filling $\nu=1$ on the square torus with $N$ bosons and $N_s=N$ magnetic flux. We label each ``layer'' by $\sigma=\uparrow,\downarrow$. The notion of ``layer'' could stand for physical layers but also any other internal degree of freedom with two components. We consider that all the particles are in the lowest Landau level and we neglect any Landau level mixing. In that case, the effective Hamiltonian is just the interaction projected onto the lowest Landau level. For our purpose, we consider the following Hamiltonian
\begin{eqnarray}
H&=&\mathcal{V}_0^{\textrm{intra}}+U_0\mathcal{V}_0^{\textrm{inter}}+U_1\mathcal{V}_1^{\textrm{inter}}\label{hamil}\\
&&-t\sum_{i=0}^{N_s-1} \left( a_{i,\uparrow}^\dagger a_{i,\downarrow} + a_{i,\downarrow}^\dagger a_{i,\uparrow}\right),\nonumber
\end{eqnarray}
where $\mathcal{V}_m^{\textrm{intra}}$ (resp. $\mathcal{V}_m^{\textrm{inter}}$) is the two-body interaction corresponding to the $m$-th Haldane's pseudo-potential\cite{Haldane-PhysRevLett.51.605} within (resp. between) layers. The last term of Eq.~(\ref{hamil}) is the inter-layer tunneling with $a_{i,\sigma}^\dagger$ (resp. $a_{i,\sigma}$) being the creation (resp. annihilation) operator for a boson in layer $\sigma$ and in the lowest Landau level orbital $i$ (with $0 \le i < N_s$). We normalize the $\mathcal{V}_m^{\textrm{intra}}$ and $\mathcal{V}_m^{\textrm{inter}}$ interaction terms such that the energy scale of the two-particle problem is of one for each of them.

The Hamiltonian of Eq.~(\ref{hamil}) possesses several symmetries. The magnetic translation invariance on the torus leads to a conserved two-dimensional momentum\cite{Haldane85-PhysRevLett.55.2095} ${\bf k}=(k_x,k_y)$ (respectively related to the relative translation and the  center of mass translation) in the Brillouin zone $k_x\in[0,2\pi),k_y\in[0,2\pi)$. Note that due to the total filling factor $\nu=1$ the reduced Brillouin zone coincides with the Brillouin zone.

The layer index can be thought as a pseudo-spin $1/2$, thus we can define the projections of the total pseudo-spin operator $\hat{{\bf S}}$ as
\begin{eqnarray}
\hat{S_x}&=&\frac{1}{2}\sum_{i=0}^{N_s - 1}\left( a_{i,\uparrow}^\dagger a_{i,\downarrow} + a_{i,\downarrow}^\dagger a_{i,\uparrow}\right)\label{SxDef}\\
\hat{S_y}&=&-\frac{\textrm{i}}{2}\sum_{i=0}^{N_s - 1}\left( a_{i,\uparrow}^\dagger a_{i,\downarrow} - a_{i,\downarrow}^\dagger a_{i,\uparrow}\right)\label{SyDef}\\
\hat{S_z}&=&\frac{1}{2}\sum_{i=0}^{N_s - 1}(a_{i,\uparrow}^\dagger a_{i,\uparrow}-a_{i,\downarrow}^\dagger a_{i,\downarrow})\label{SzDef}
\end{eqnarray}
In this language, the tunneling term of Eq.~(\ref{hamil}) is the analogue of a Zeeman term along the fictitious $x$ axis. Thus a large tunneling amplitude $t$ has the effect of polarizing the system along the $x$ axis. At zero interlayer tunneling $t=0$, $\hat{S_z}$ is a good quantum number with eigenvalues $S_z=\frac{1}{2}(N_\uparrow - N_\downarrow)$, where $N_\uparrow$ (resp. $N_\downarrow$) is the particle number in the up (resp. down) layer. Furthermore, if $U_0=1$ and $t=0$, the Hamiltonian of Eq.~(\ref{hamil}) exhibits a full $SU(2)$ symmetry {\it irrespective of } $U_1$, since bosons with identical spin cannot feel odd Haldane's pseudo-potentials. In that case, not only $\hat{S_z}$ but also the total pseudospin $\hat{{\bf S}}^2$ are conserved quantities.

If we set $U_0=U_1=t=0$, Eq.~(\ref{hamil}) becomes the model Hamiltonian for the $(220)$ Halperin state which is just two decoupled copies of the $\nu=\frac{1}{2}$ Laughlin state. Indeed they are the densest zero-energy eigenstates of the $\mathcal{V}_0^{\textrm{intra}}$ interaction, i.e. the hardcore interaction projected onto the lowest Landau level. The $(220)$ Halperin state falls in the $S_z = 0$ sector and is four-fold degenerate on the torus geometry. These four states respectively carry the momentum quantum numbers ${\bf k}=(0,0)$, $(\pi,0)$, $(0,\pi)$, and $(\pi,\pi)$. The topological degeneracy is the most practical signature of topological order since it can be directly extracted from the energy spectrum. We will therefore use it extensively in this paper to distinguish different phases.

\subsection{The coupled Moore-Read state}\label{sec:coupledMR}
Among the other possible phases that might emerge in a bosonic bilayer at $\nu=1$, Ref.~\onlinecite{Hormozi-PhysRevLett.108.256809} introduced the coupled MR (cMR) state. The physical picture of the cMR can be thought as two chiral $p$-wave superconductors of composite fermions with a tunneling of Cooper pairs. In the $S_z=0$ sector and on the plane geometry the wave function of the cMR state possesses a simple and elegant expression

\begin{eqnarray}
\Psi_{\rm cMR}&=&{\rm Pf}\left(\frac{1}{z^\uparrow_i-z^\uparrow_j}\right) \; {\rm Pf}\left(\frac{1}{z^\downarrow_i-z^\downarrow_j}\right)\nonumber\\
&&\times \prod_{\sigma=\uparrow,\downarrow}\prod_{i<j}\left(z^\sigma_i-z^\sigma_j\right)\; \prod_{i,j}\left(z^\uparrow_i-z^\downarrow_j\right).
\label{coupledMR}
\end{eqnarray}
Here the $z_i^\uparrow$'s (resp.  $z_i^\downarrow$'s) are the particle complex coordinate in the upper (resp. lower) layer. This wave function is the exact densest zero energy state of the following model Hamiltonian\cite{Moller-PhysRevB.90.235101} (once projected onto the lowest Landau level)

\begin{eqnarray}
H_{3-2}&=&\sum_{\sigma=\uparrow,\downarrow}\sum_{i<j<k}\delta^{(2)}\left(z^\sigma_i-z^\sigma_j\right)\delta^{(2)}\left(z^\sigma_j-z^\sigma_k\right)\nonumber\\
&& + \sum_{i,j}\delta^{(2)}\left(z^\uparrow_i-z^\downarrow_j\right).\label{hamiltonian32}
\end{eqnarray}
This Hamiltonian has two types of interaction: a three-body hardcore interaction within each layer and a two-body hardcore interaction between layers, i.e., a $\mathcal{V}_0^{\textrm{inter}}$ term. The degeneracy of the cMR state is richer on the torus than on the plane geometry. Indeed, the number of zero-energy states of the Hamiltonian Eq.~(\ref{hamiltonian32}) at filling factor $\nu=1$ is the following:
\begin{itemize}
\item If $N = 4m$, $m\in \mathbb Z$: 3 zero-energy states in the even $S_z$ sectors, respectively carrying the momenta $\mathbf{k} = (0,0)$, $(0,\pi)$ and $(\pi,0)$. One zero-energy state at $(\pi,\pi)$ in the odd $S_z$ sectors.
\item If $N=4m+2$, $m\in \mathbb Z$: 3 zero-energy states in the odd $S_z$ sectors, respectively carrying the momenta $\mathbf{k} =(0,0)$, $(0,\pi)$ and  $(\pi,0)$. One zero-energy state at $(\pi,\pi)$ in the even $S_z$ sectors.
\item If $N = 2m + 1$, $m \in \mathbb Z$: only one zero-energy state in the $S_z=\pm N/2$ sector at momentum $(0,0)$
\end{itemize}
Note that in the fully polarized sector $S_z=\pm N/2$, the ground state is just the usual single-component MR state.
Thus the total degeneracy is $2N+3$ when $N$ is even and $2$ when $N$ is odd. The extensive degeneracy stresses the gapless nature of the Hamiltonian of Eq.~(\ref{hamiltonian32}).
Nevertheless, Ref.~\onlinecite{Moller-PhysRevB.90.235101} has considered the addition of some Josephson coupling that could be written as
\begin{eqnarray}
H_{\rm J}&=&t_{\rm J} \mathcal{V}_0^{\uparrow\uparrow;\downarrow\downarrow}\; +\; \rm{h.c.},\label{josephson}
\end{eqnarray}
where $\mathcal{V}_0^{\uparrow\uparrow;\downarrow\downarrow}$ is a $0$-th Haldane's pseudo-potential coupling two spin-up to two spin-down bosons. They showed that a small $t_{\rm J}$ between the two layers lifts the extensive degeneracy and opens a gap of order $t_{\rm J}$. This gapped phase has the same nature as the Halperin $(220)$ state. Interestingly, the ground state state of the Hamiltonian Eq.~(\ref{hamiltonian32}) at $\nu=1$ in the presence of an infinitesimal (but non-zero) tunneling is nothing but the Halperin $(220)$ state in a rotated spin basis (with a $\pi/2$ rotation around the $y$ spin axis) and up to finite-size corrections that quickly vanish.

\begin{figure*}
\centerline{\includegraphics[width=0.8\linewidth]{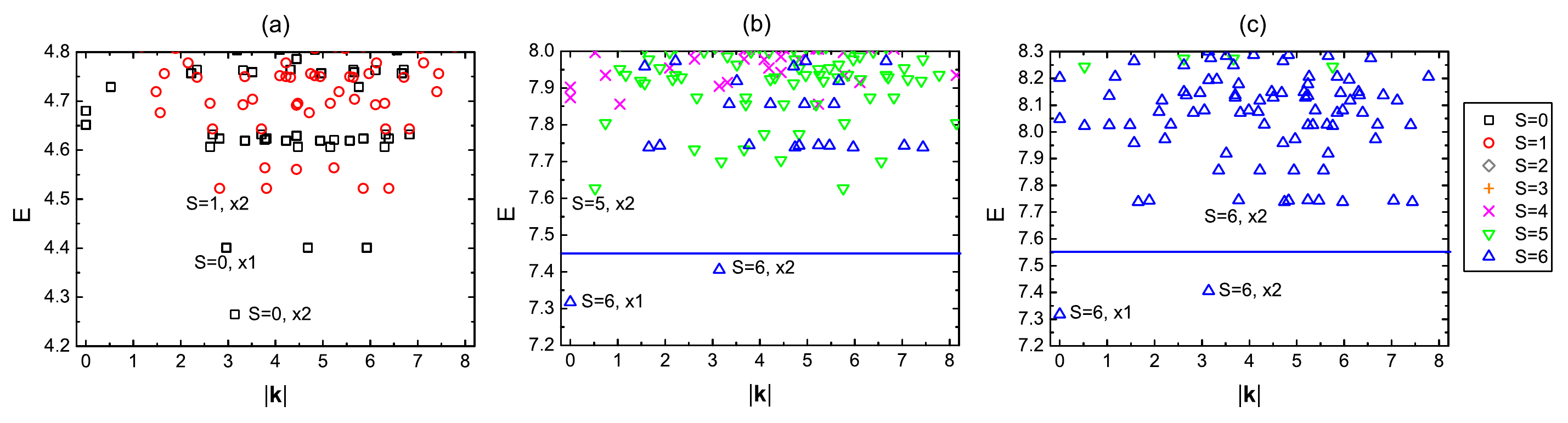}}
\caption{(Color online) The energy spectra of $N=12$ bosons as a function of momentum $|{\bf k}|=\sqrt{k_x^2+k_y^2}$ at $U_0 = 1$ and (a) $U_1=0$, (b) $U_1=0.4$ and (c) $U_1=1$. The spin eigenvalues $S$ of some low-lying states as well as their degeneracy (the spin degeneracy has been excluded) are indicated. One can see that the ferromagnetic levels dominate the low-energy spectrum from the bottom with the increase of $U_1$. The states below the blue lines are the three quasi-degenerate ferromagnetic MR states appearing at ${\bf k}=(0,0)$, $(\pi,0)$, and $(0,\pi)$.}
\label{FM}
\end{figure*}

\section{Moore-Read state in the fully polarized regime}\label{sec:MR}

Using the analogy between the spin and the layer index, the situation where all the particles are in one layer is called the fully polarized regime, i.e., $S_z=\pm N/2$. This situation is relevant to our bilayer system in different cases. The system can become polarized in a rotated basis due to a strong tunneling term, as mentioned in Sec.~\ref{sec:model}. In the absence of tunneling, and for an $SU(2)$-symmetric interaction, the system can again be dominated by the single-layer physics if the ground state is ferromagnetic ($S = N/2$).

In the fully polarized regime, our system is identical to a problem of single-layer bosons with the contact interaction $\mathcal{V}_0$ (dropping the $\textrm{intra}$ label) at filling factor $\nu=1$. Previous studies\cite{Cooper-PhysRevLett.87.120405,regnault-PhysRevLett.91.030402,Chang-PhysRevA.72.013611,PhysRevB.69.235309,PhysRevB.76.235324} have shown strong evidence that the emerging phase would be described by the MR state when the number of particles $N$ is even. Here we provide more abundant numerical data (Fig.~\ref{polarized}) to establish the stability of the MR phase for both even and odd $N$ in the fully polarized regime, as the basis of our discussion of the bilayer phase diagram.

When $N$ is even, a straightforward hallmark of the bosonic MR phase is the three-fold ground-state topological degeneracy on the torus geometry with one state in each momentum sector ${\bf k}=(0,0)$, $(\pi,0)$ and $(0,\pi)$. This degeneracy is exact when we consider the three-body contact interaction for which the MR states are the exact densest zero-energy eigenstates. For more realistic two-body interactions such as the two-body contact interaction, this degeneracy is expected to be recovered only in the thermodynamic limit except if it is enforced by a discrete symmetry. Indeed using a square torus implies a $C_4$ symmetry, such that the energy levels are identical in the $(\pi,0)$ and $(0,\pi)$ momentum sectors. In our numerical data, we indeed observe three almost degenerate states in the expected ${\bf k}$ sectors. In Fig.~\ref{polarized}(a), we show the energy gap $\Delta_{\rm MR}$ between the highest energy state of the three-fold low-energy manifold associated to the MR state and the first excited state (irrespective of its momentum) as well as the low-energy manifold energy splitting $\delta_{\rm MR}$. The data are given for various system sizes, up to $N=20$ bosons. The topological degeneracy is not that clear for small system sizes due to the strong finite-size effects, but it is greatly improved for $N\geq12$. We can see that the energy gap $\Delta_{\rm MR}$ tends towards a constant while the ground-state splitting $\delta_{\rm MR}$ starts to vanish, suggesting a recovery of the exact degeneracy and a finite gap in the thermodynamic limit. Note that $\Delta_{\rm MR}$ shows more finite size effects than the equivalent quantity obtained with the model three-body interaction\cite{Repellin-PhysRevB.92.115128}. We then compute the ground-state overlap (defined as the square norm of the scalar product) with the exact MR state obtained by diagonalizing the three-body contact interaction or by projective construction on the torus\cite{Repellin-PhysRevB.92.115128} (especially for the largest systems). While the overlap unavoidably decreases with the system size, it is still convincingly high even for the largest samples (see Fig.~\ref{polarized}(b)).

The MR state usually implies an even number of particles. However compared to geometries with zero genus (such as the disk or the sphere), the torus geometry allows the existence of a single MR state at filling factor $\nu=1$ with an odd number of particles in the ${\bf k}=(0,0)$ sector. Thus, we have also computed the energy gap and the ground-state overlap for the odd $N$ case. These data are shown together with the even $N$ case in Fig.~\ref{polarized}. The calculations have been done up to $N=19$. As can be observed, the gap has slightly more important finite-size effects than the even $N$ case (but the overlaps are a bit larger). Still these results convincingly support the emergence of the MR phase in the odd $N$ sector. It is expected that the energy gap for both even $N$ and odd $N$ will converge to the same value in the thermodynamic limit. Indeed each $N$ sector should exhibit on the torus two types of neutral excitation modes: a magneto-roton mode\cite{Yang-PhysRevLett.108.256807,moller-PhysRevLett.107.036803} and a neutral fermion mode\cite{moller-PhysRevLett.107.036803, bonderson-PhysRevLett.106.186802,Repellin-PhysRevB.92.115128}. Their respective dispersion relation should not depend on the particle number parity (as can be seen in Ref.~\onlinecite{Repellin-PhysRevB.92.115128} for the three-body model interaction). For the two-body contact interaction, our results are compatible with such a property up to more important finite-size effects.

Beyond energetics and overlap calculations, we can use the particle-cut entanglement spectrum (PES) \cite{PhysRevLett.106.100405} to probe the topological order of the phase. We divide the whole system into two parts $A$ and $B$ with $N_{A}$ and $N_{B}=N-N_A$ bosons respectively. The reduced density matrix $\rho_A$ is obtained by tracing out the $B$ part of the density matrix ($\rho_A=\textrm{Tr}_{B}\rho$). For an even number of particles, the density matrix of the ground-state manifold writes $\rho=\frac{1}{3} \sum^3_{\alpha=1}|\Psi_\alpha\rangle\langle\Psi_\alpha|$, where $|\Psi_\alpha\rangle$ represents the $\alpha$-th ground state [for an odd number of particles $\rho$ is just the projector on the single ground state at ${\bf k}=(0,0)$]. Diagonalizing the reduced density matrix gives access to the PES, whose levels are $\xi_i = -\ln\lambda_i$ ($\lambda_i$ is the $i$-th eigenvalue of the reduced density matrix). When the low-energy manifold is in the MR phase, we expect to observe an entanglement gap separating the low-lying levels from the high non-universal levels. We also expect the number of low-lying level $d$ to be the same as the number of MR quasihole excitations in a system with $N_A$ particles and the same number of orbitals. This number can be predicted by the generalized exclusion rule\cite{PhysRevLett.100.246802} of the MR state, i.e., no more than two bosons in two consecutive orbitals. The entanglement gap between the low-lying levels and the first excited level is then defined as $\Delta_{\xi}\equiv \xi_{d+1}-\xi_{d}$ (the PES levels are sorted in increasing order). We have performed this calculation for systems with both parities of $N$. We find that the entanglement gap is indeed finite, and does not vanish with the increase of the system size [Fig.~\ref{polarized}(c)], meaning the ground state has the same quasihole excitation properties as the MR state.

\section{Inter-layer interaction effect}\label{sec:phasediagraminter}

In this section, we assume zero inter-layer tunneling $t=0$, and study the phase diagram in the $U_0-U_1$ space.

\subsection{The $SU(2)$ symmetric regime}\label{sec:su2inter}

We start our exploration by focusing on the $SU(2)$ invariant line at $U_0=1$ as discussed in Sec.~\ref{sec:model}. In that case, each eigenstate of the Hamiltonian (\ref{hamil}) can be labeled by both $S$ and $S_z$, where $0\leq S\leq N/2$ and $-S\leq S_z\leq S$. We are particularly interested in those ferromagnetic states with maximal $S=N/2$. The spatial part of theses eigenstates coincides with the one of the fully polarized regime in the $S=S_z=\pm N/2$ sector. As shown by the extensive numerical study of the previous section, the fully polarized system hosts a robust MR phase. Thus the MR phase is guaranteed to emerge as the ground state manifold of the ferromagnetic states. If these ferromagnetic states dominate the low-lying spectrum of the Hamiltonian (\ref{hamil}) then the low-energy physics of the system will be driven by the single-layer picture and captured by the MR phase.

\begin{figure}
\centerline{\includegraphics[width=0.6\linewidth]{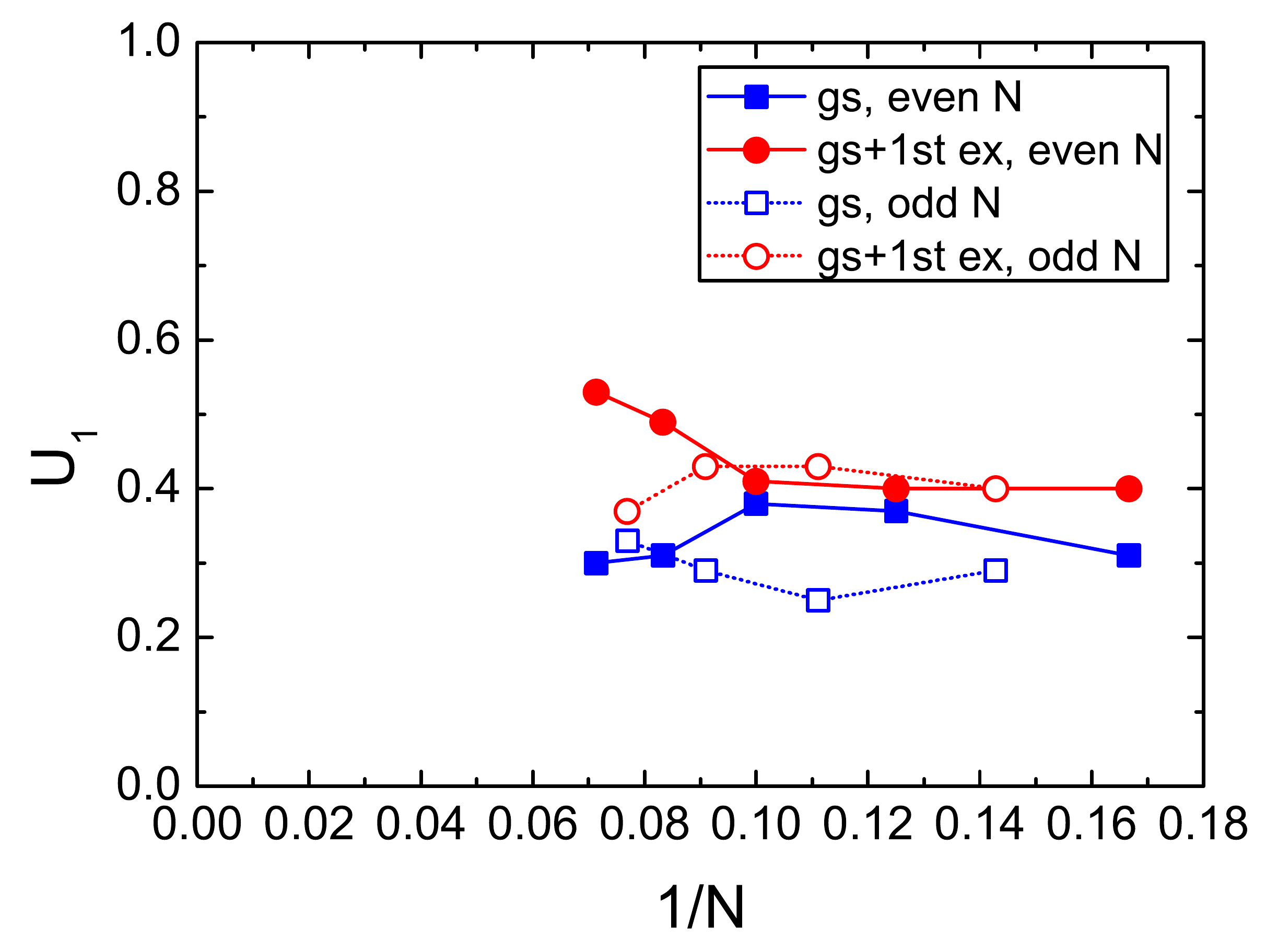}}
\caption{(Color online) The critical value of $U_1$ at which the ferromagnetic MR manifold and its first ferromagnetic excitation dominate the low-energy spectrum for even $N$ (solid line) and odd $N$ (dotted line). The data from $N=6$ to $N=14$ are included.}
\label{U1}
\end{figure}

We track the evolution of the low-lying spectrum of the Hamiltonian (\ref{hamil}) with $U_1$. A typical example of the low energy spectrum for an even number of particles is shown in Fig.~\ref{FM}. At $U_1=0$, the low-lying levels have small $S$ values such as $S=0$ and $S=1$ [Fig.~\ref{FM}(a)], and the ferromagnetic states are still high in energy. However, the energy levels with $S<N/2$ ascend with the increase of $U_1$. When $U_1$ is increased up to a critical value, ferromagnetic states [Figs.~\ref{FM}(b)] begin to have a lower energy than other $S<N/2$ levels. Beyond this critical point the low energy physics is dominated by the (ferromagnetic) MR states. We have studied this critical value of $U_1$ for various system sizes with both even and odd $N$ (up to $N=14$, see Fig.~\ref{U1}). An extrapolation to $1/N\rightarrow0$ suggests $U_1=U^c_1\approx0.2-0.3$ in the thermodynamic limit.

The situation of the excited states is more complex. Indeed, in the regime where the ferromagnetic MR manifold has the lowest energy, the low energy excited states may still have $S<N/2$, corresponding to spinful excitations. However, if we further increase $U_1$ ($U_1\approx0.6-0.7$ in the thermodynamic limit), not only the ground manifold but also the first excited level belong to the ferromagnetic states [Figs.~\ref{FM}(c)]. In that case, the energy gap is the same as the one of the fully polarized regime. Note that the spinful excitations of the MR states have been studied both numerically\cite{Wojs-PhysRevLett.104.086801} and analytically\cite{Romers-NJP-045013} in the context of the fermionic $\nu=\frac{5}{2}$ FQHE. But to our knowledge, no study has considered the bosonic case.

\begin{figure}
\centerline{\includegraphics[width=0.7\linewidth]{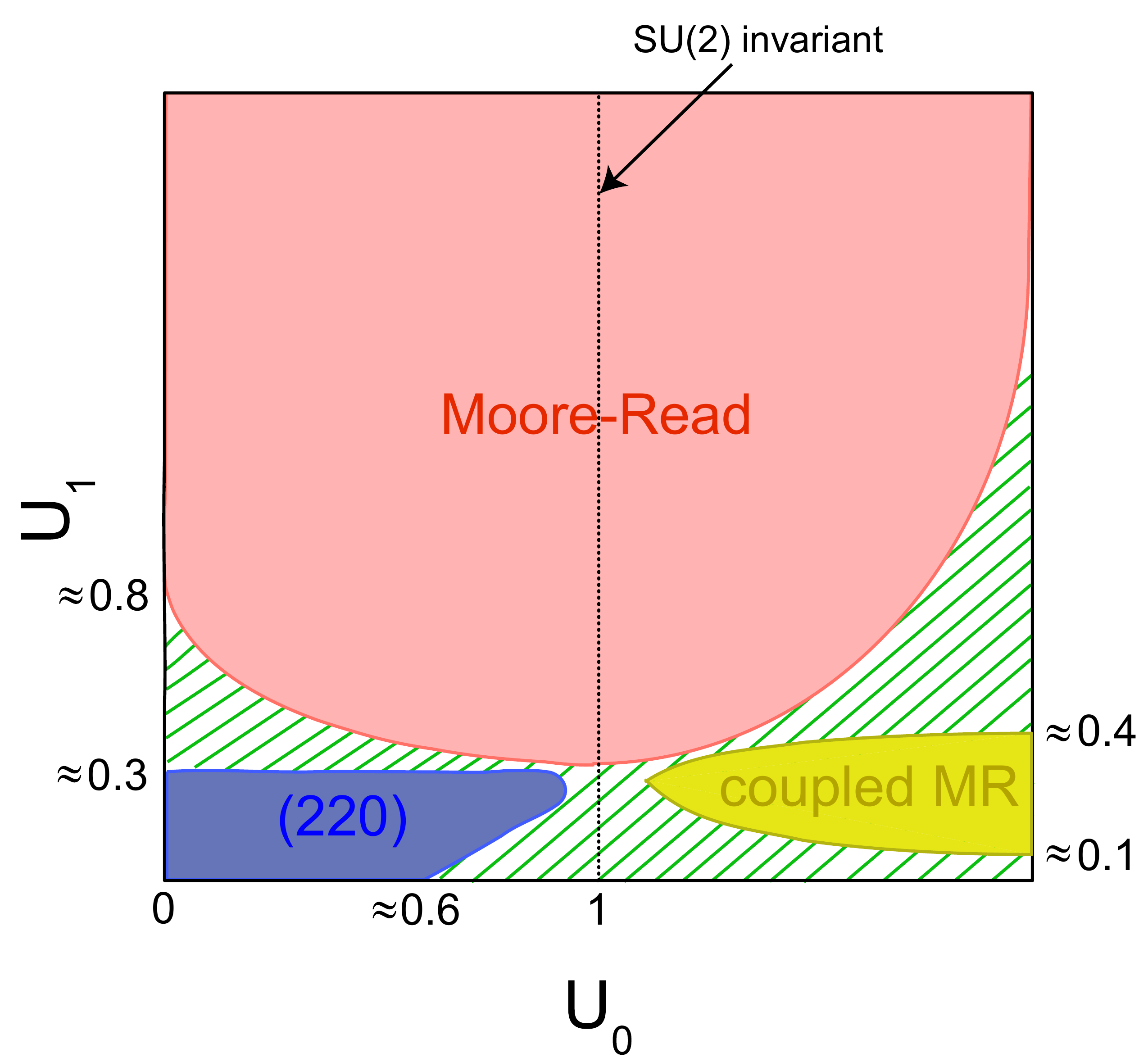}}
\caption{(Color online) A schematic $U_0-U_1$ phase diagram for even $N$ in the $S_z=0$ sector.
As discussed in the text, we observe three phases: $(220)$ Halperin, Moore-Read, and coupled Moore-Read state.
The rough ranges of these phases are indicated in the figure. The shadowed areas between the three phases are
transition regions, whose properties are difficult to be identified based on our present numerical data and could be compressible. It is also difficult to tell if a direct transition between the $(220)$ Halperin and MR (or between the cMR and MR) could occur.}
\label{phasediagram}
\end{figure}

\begin{figure*}
\centerline{\includegraphics[width=\linewidth]{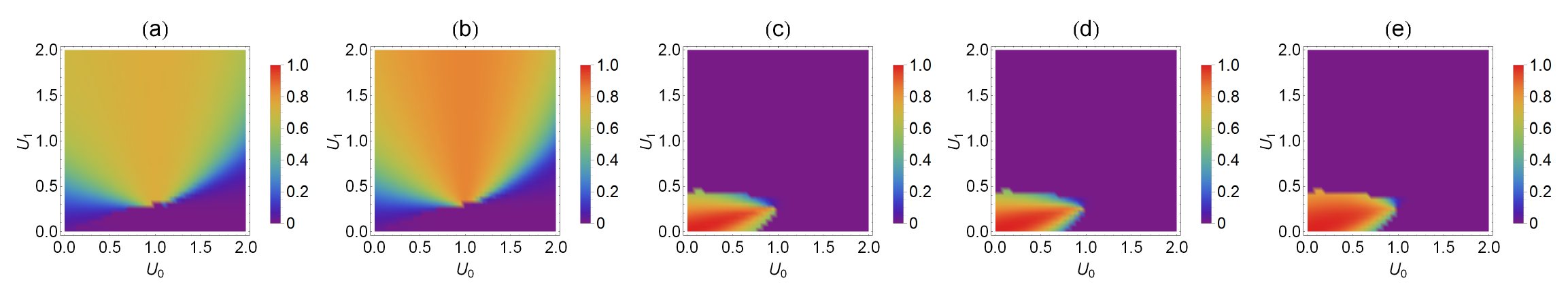}}
\caption{(Color online) Nature of the low-energy manifold as a function of $U_0$ and $U_1$ for a system of $N=12$ bosons in the $S_z=0$ sector. (a)-(b) The ground-state overlap with the exact MR state in (a) ${\bf k}=(0,0)$ and (b) ${\bf k}=(\pi,0)$ sector. (c)-(e) The ground-state overlap with the exact $(220)$ state in (c) ${\bf k}=(0,0)$, (d) ${\bf k}=(\pi,0)$ and (e) ${\bf k}=(\pi,\pi)$ sector. The overlap in the ${\bf k}=(0,\pi)$ sector is identical to that in the ${\bf k}=(\pi,0)$ sector due to the $C_4$ symmetry of the square torus.}
\label{overlapv0v1}
\end{figure*}

\begin{figure}
\centerline{\includegraphics[width=\linewidth]{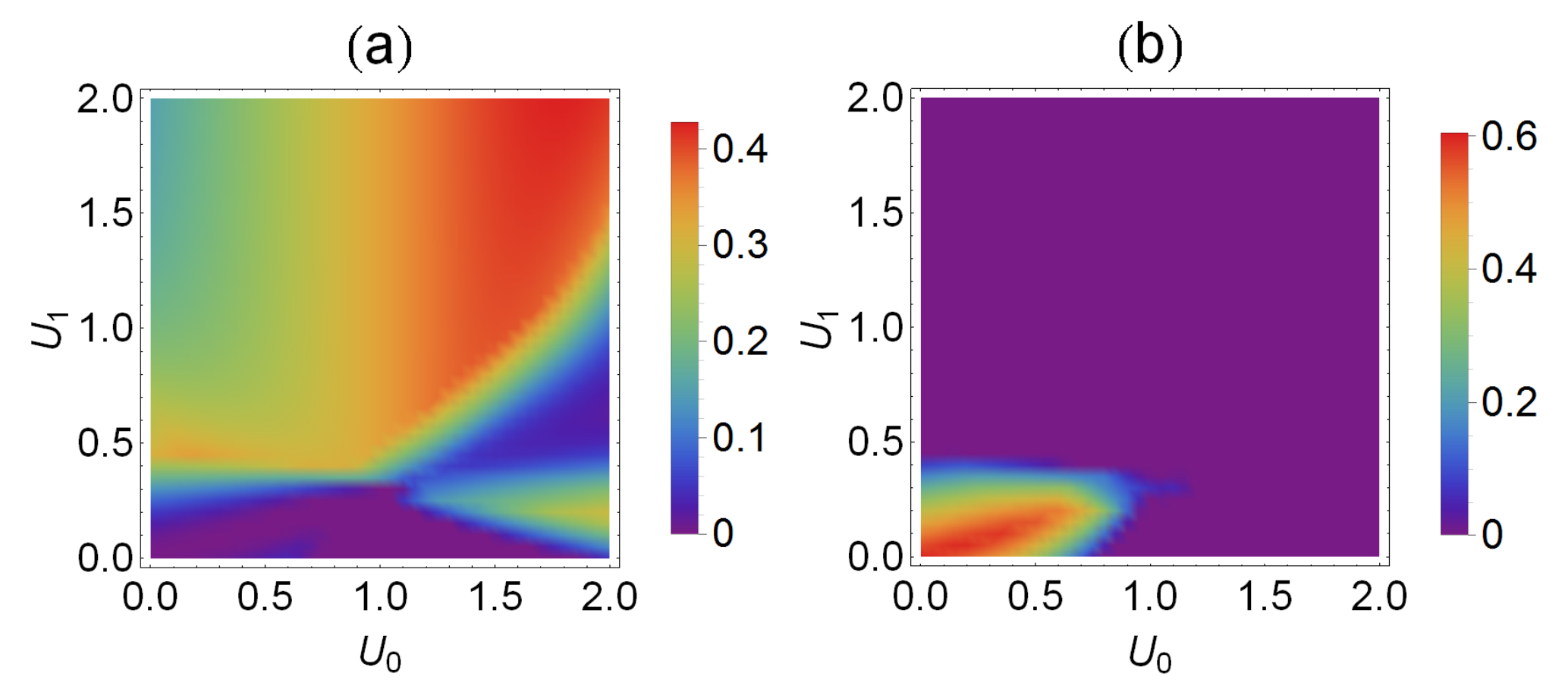}}
\caption{(Color online) The energy gap relative to (a) the MR manifold and (b) the $(220)$ manifold for $N=12$ bosons in the $S_z=0$ sector.}
\label{gap12}
\end{figure}

\begin{figure}
\centerline{\includegraphics[width=\linewidth]{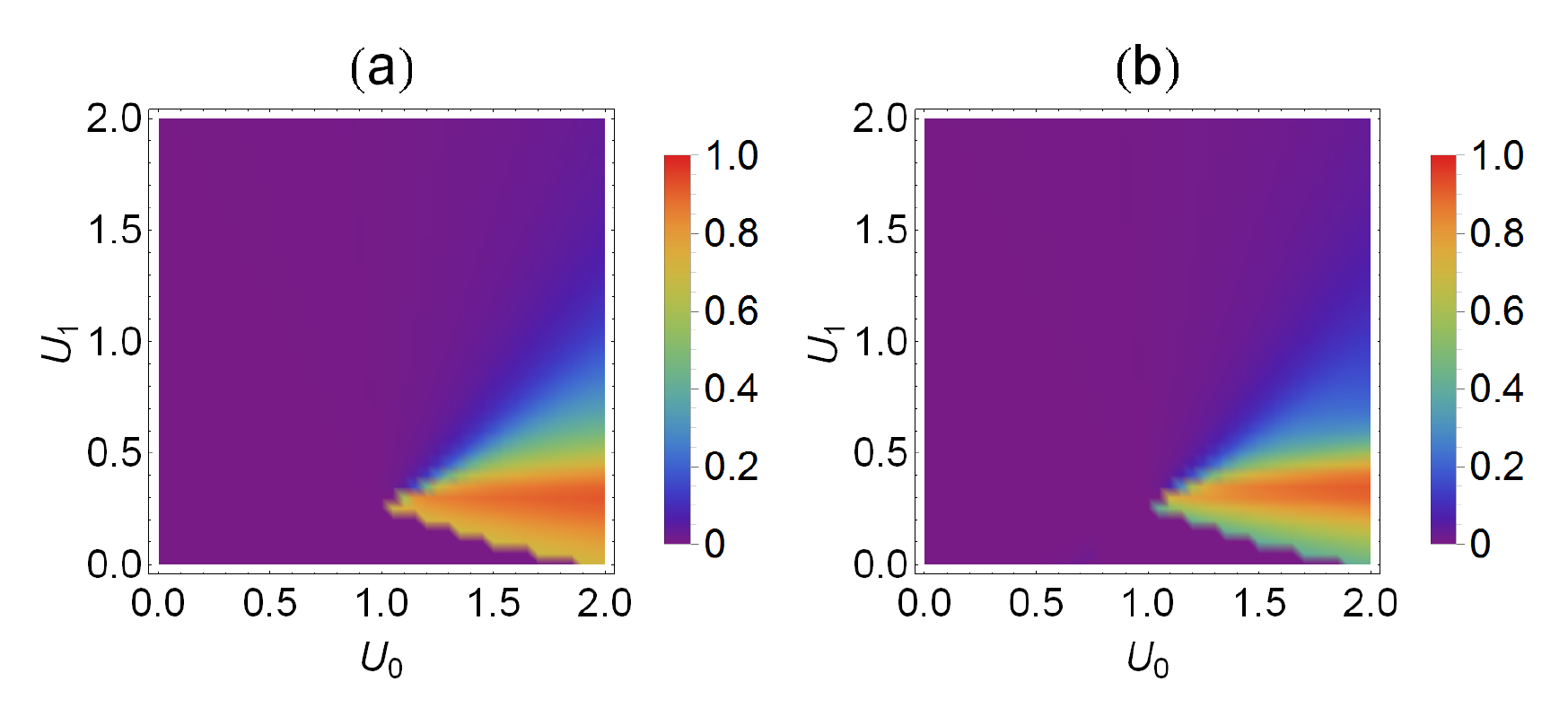}}
\caption{(Color online) The ground-state overlap with the exact cMR state for $N=12$ bosons in the $S_z=0$ sector at (a) ${\bf k}=(0,0)$and (b) ${\bf k}=(\pi,0)$. In the $U_0-U_1$ region shown here, the overlap in both momentum sectors is maximum around the point $U_0\approx 2.0$ and $U_1 \approx 0.35$ (up to $0.91$).}
\label{cMR12}
\end{figure}

\subsection{The $U_0-U_1$ phase diagram}\label{sec:u0u1inter}

The $SU(2)$ invariance is broken if $U_0$ is tuned away from $1$. Consequently the total pseudo-spin $S$ is no longer a good quantum number. We first focus on the $S_z=0$ (resp. $S_z=\frac{1}{2}$) sector for even (resp. odd) number of particles to explore the phase diagram which is summarized in Fig.~\ref{phasediagram}. For $N$ even, there are two obvious candidate phases that we might consider: the MR state and the Halperin $(220)$ state. The former should at least appear along the $SU(2)$ invariant line $U_0=1$ and beyond the critical value $U_1 > U^c_1$. The later should be present around the point $U_0=U_1=0$ for which the $(220)$ state is the exact ground state. We also know that this latest state cannot be a correct description of the low energy physics at $U_0=1$ irrespective of $U_1$ since it explicitly breaks the $SU(2)$ symmetry.

We compute the ground-state overlap with the respective exact model states and the energy gap above the MR or the $(220)$ state to determine the range of these two phases. The overlap is set to $0$ if the whole manifold of the ground states is not in the same momentum sector as the model states. The exact MR states in the $S_z=0$ sector are generated by consecutively applying the spin ladder operator $S^-=\sum_{i=0}^{N_s-1} a_{i,\downarrow}^\dagger a_{i,\uparrow}$ on the fully polarized version in the $S_z=N/2$ sector. When computing the energy gap above the $(220)$ (resp. MR) state, we pick up the lowest four (resp. three) energy states irrespective of their momenta. If they are in the same ${\bf k}$ sectors as the $(220)$ (resp. MR) state, we define the gap as the difference between the highest energy in this manifold and the first excited level above it, otherwise the gap is set to $0$. We find that the numerical data are qualitatively identical for $N=8$, $N=10$ and $N=12$, implying the finite-size effects on the phase boundaries are small. Note that this relative independence to the system size has also been observed for a lattice realization of this bilayer\cite{PhysRevB.90.245401} along the $U_1=0$ line.

In Figs.~\ref{overlapv0v1} and \ref{gap12}, we provide a typical example of numerical results for a system with $N=12$. As already mentioned in Sec.~\ref{sec:model}, the $(220)$ state is the exact ground state of the Hamiltonian at $U_0=U_1=0$. We thus find the $(220)$ phase around this point. It becomes unstable with the moderate increase of either $U_0$ or $U_1$. In the $U_1$ direction, the $(220)$ phase collapses at $U_1\approx0.3$ with almost no dependence on $U_0$. A smooth transition to the MR phase occurs in ${\bf k}=(0,0)$, $(\pi,0)$ and $(0,\pi)$ sectors. Note that in finite size the MR and $(220)$ have a small but non-zero overlap [for $N=12$, these overlaps are $\approx 0.027$ for the $(0,0)$ momentum sector and $\approx 0.023$ for the $(0,\pi)$ and $(\pi,0)$ momentum sectors for an Hilbert space dimension of $\simeq 10^6$]. In the ${\bf k}=(\pi,\pi)$ sector, there is a clear level crossing signature because the overlap suddenly drops to $0$ [Fig.~\ref{overlapv0v1}(e)], implying that the $(220)$ state in this ${\bf k}$ sector goes up and finally mixes in the excited states. In the $U_0$ direction, the $(220)$ phase can survive up to $U_0\approx0.6$ for $U_1=0$ (consistent with the result in Ref.~\onlinecite{PhysRevB.90.245401}), and a larger value of $U_0\approx0.8$ for $U_1\approx0.3$. Beyond this point, a transition to another phase occurs. The critical $U_1$ value for the MR phase is about $0.3$ around $U_0=1$, but increases fast when $U_0$ is tuned away from $1$. At the largest $U_0$ that we study ($U_0=2$), the system enters the MR phase at $U_1\approx1.5$.

Looking more carefully at the Fig.~\ref{gap12}(a), we observed another region around $U_0=2$, $U_1=0.3$, where there is a reentrant energy gap above the ${\bf k}=(0,0)$, $(\pi,0)$ and $(0,\pi)$ sectors. By examining the energy spectra in all $S_z$ sectors, we observe a low-energy degeneracy pattern consistent with the coupled Moore-Read state introduced in Sec.~\ref{sec:coupledMR}. We then compute the ground-state overlap with the model cMR state. Indeed, the overlap becomes high when the gap reopens (Fig.~\ref{cMR12}), confirming the presence of the cMR phase in this region. Note that this phase collapses when $U_1\gtrsim0.4$ with almost no dependence on $U_0$.

\begin{figure}
\centerline{\includegraphics[width=0.6\linewidth]{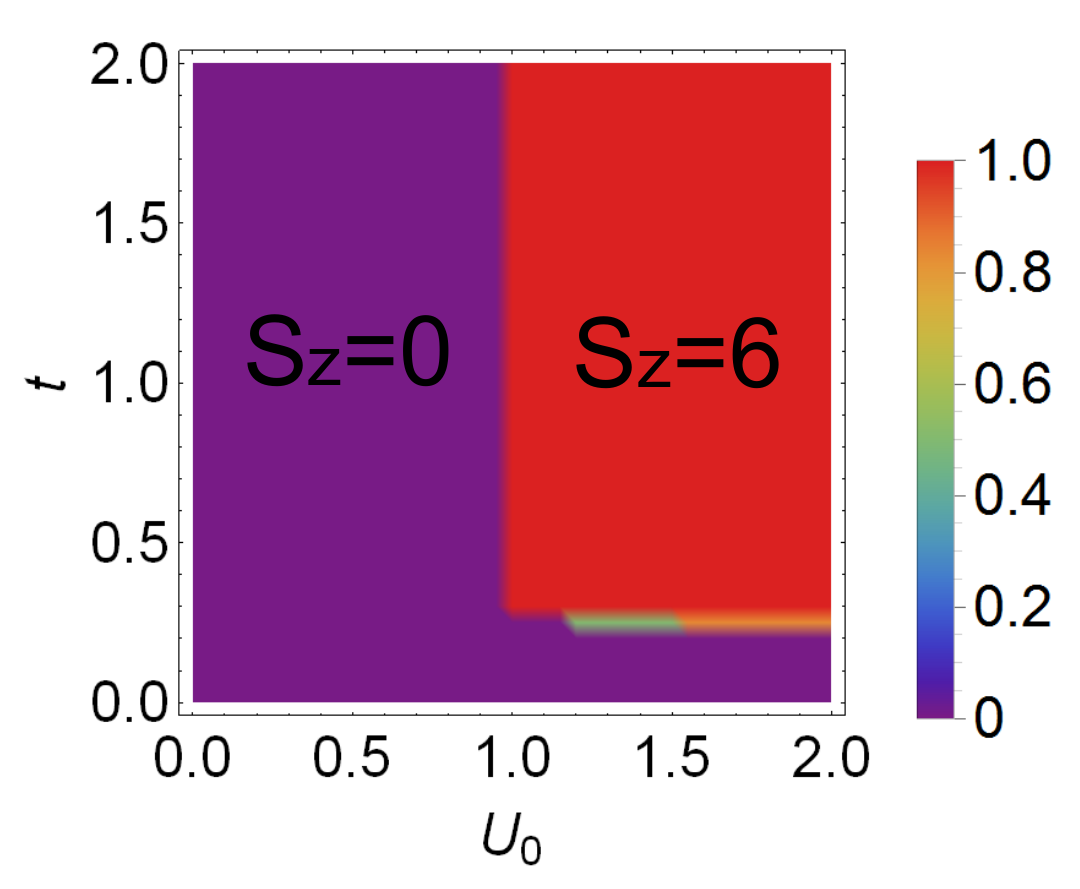}}
\caption{(Color online) The polarization ${\cal P}_z$ along the $z$ spin axis for the absolute ground state and $N=12$ bosons. $S_z$ being a good quantum number, the ${\cal P}_z$ has only a discrete number of values. The purple region is unpolarized ($S_z=0$) while the red region is fully polarized ($S_z=N/2=6$). This latest corresponds to $U_0 > 1.0$ and $U_1\gtrsim0.3$. }
\label{Pz12}
\end{figure}

Beyond the $S_z=0$ sector, we can wonder if the system exhibits some regions with a spontaneous polarization in part of the phase diagram. For that purpose, we compute the $z$ polarization ${\cal P}_z= \langle \hat{S_z} \rangle / (N/2)$ of the absolute ground state (without focusing on a specific quantum number sector). Note that here we simply have $\langle \hat{S_z} \rangle=S_z$. As can be observed in Fig.~\ref{Pz12}, the system is fully polarized when $U_0>1$ and $U_1>U^c_1$. Clearly the low energy physics in that region is governed by the MR state. In the rest of diagram, the system is unpolarized leading to the different phases described previously for $S_z=0$.

\begin{figure}
\centerline{\includegraphics[width=\linewidth]{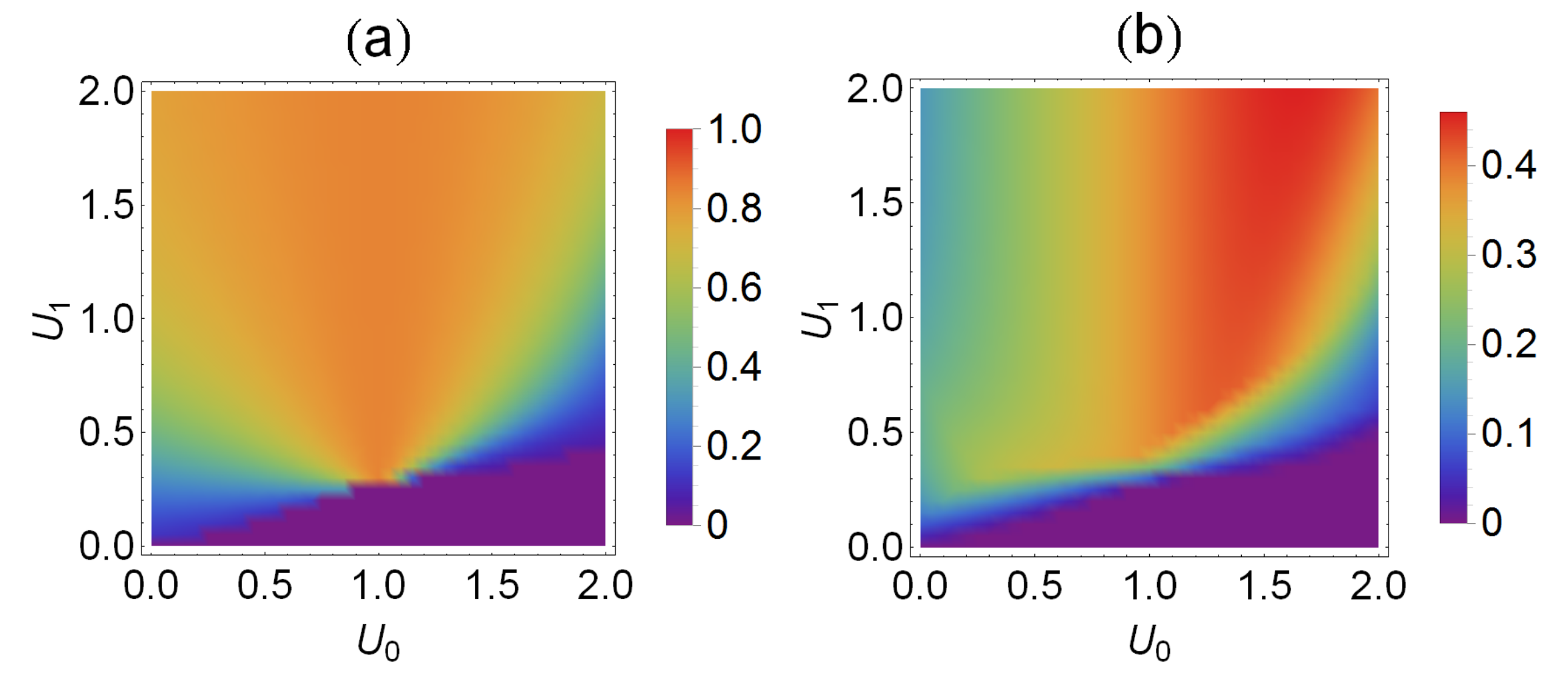}}
\caption{(Color online) (a) The ground-state overlap with the exact MR state in the ${\bf k}=(0,0)$ sector. (b) The energy gap relative to the MR manifold. Here we consider $N=11$ bosons in the $S_z=\frac{1}{2}$ sector.}
\label{np11}
\end{figure}

We now turn to the odd $N$ sector. The Halperin $(220)$ state cannot be realized in that case since it requires an equal number of particles in each layer. Still there is a single MR state in the ${\bf k}=(0,0)$ sector as mentioned in Sec.~\ref{sec:MR}. Fig.~\ref{np11} gives both the overlap between the ground state of the bilayer system and the odd particle MR state, and the gap relative to the MR state (in the $S_z=\frac{1}{2}$ sector). As shown, the situation is almost identical to the $N$ even case. This is also true for the $z$ polarization (not shown here).

We can summarize the different phases observed in this set-up via the schematic phase diagram given in Fig.~\ref{phasediagram}. We have clear evidence for three phases related to the following states: the Halperin $(220)$ state, the MR state and the cMR state. Unfortunately, exact diagonalizations (ED) do not allow to probe the transition between these phases. For instance, we cannot rule out a compressible phase that would lie between e.g. the Halperin $(220)$ phase and the MR phase. There is also a small region around the $SU(2)$ symmetric point $U_0=1$ and $U_1=0$ with no clear gap structure [see Fig.~\ref{FM}(a)] and its nature might be compressible. Actually, some clues of composite fermion sea were observed in Ref.~\onlinecite{PhysRevA.91.063623} at this point. Moreover, the results obtained by Ref.~\onlinecite{Wu-PhysRevB.87.245123} allows to rule out a candidate such as the Jain spin singlet\cite{Moran-PhysRevB.85.245307}.

\subsection{A deeper look at the coupled MR state}

The ground-state overlap with the exact cMR state is maximal close to $U_0=2.0$ and $U_1=0.35$ for $N=12$ (see Fig.~\ref{cMR12}). While the overlaps are high there, we would like to look for additional signature of the cMR state, to probe the topological order in this region of the phase diagram. Here we consider several system sizes from $N=8$ to $N=14$ and we will focus on the $U_0=2.0$ and $U_1=0.35$ as a typical candidate of the phase near this point.

\begin{figure}
\centerline{\includegraphics[width=0.85\linewidth]{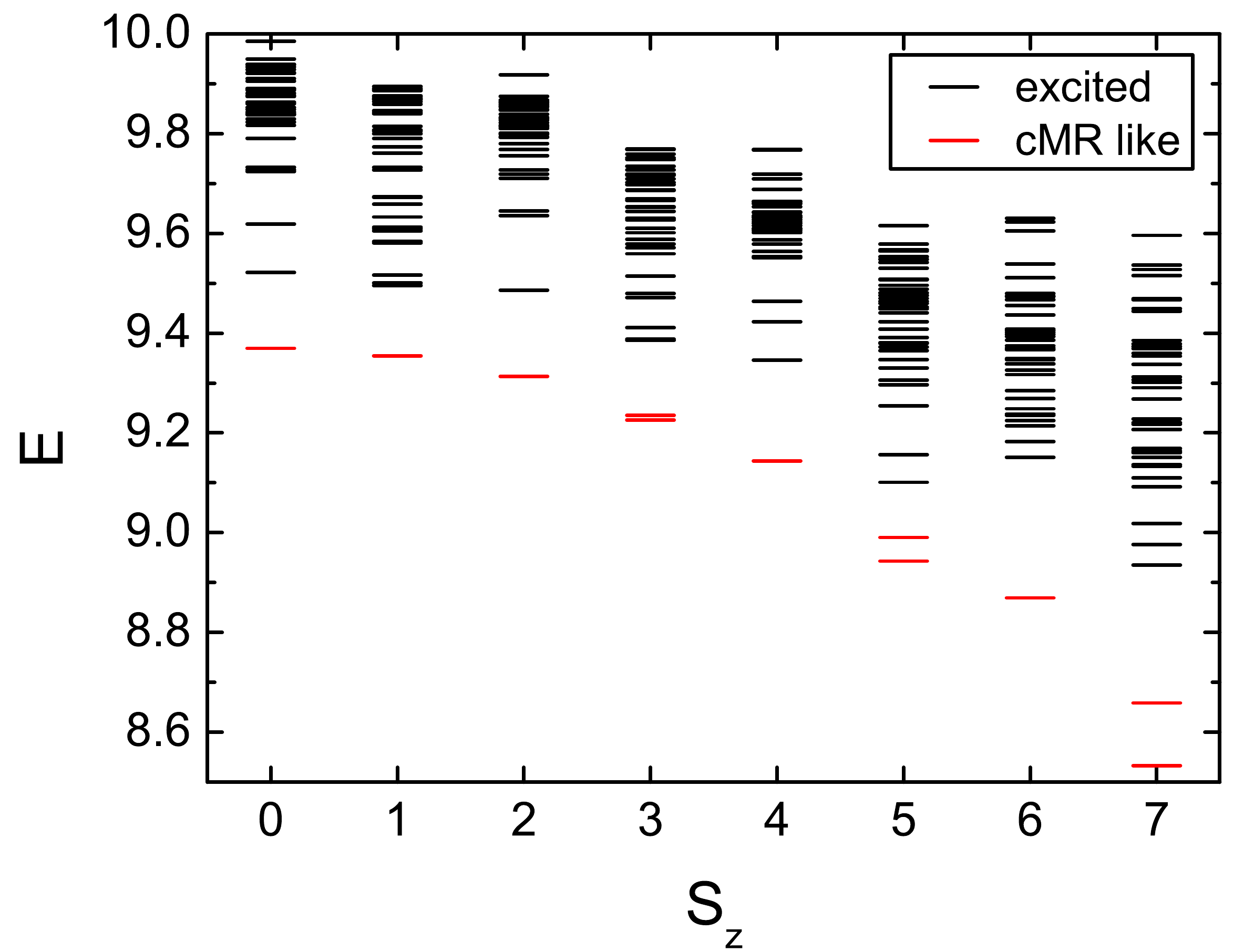}}
\caption{(Color online) The low energy spectrum as a function of $S_z$ for $N=14$ bosons at $U_0=2.0$ and $U_1=0.35$. The red symbols correspond to the cMR like states. We see the alternation of the degeneracy : a unique state when $S_z$ is even, threefold (with one exact degeneracy due to the $C_4$ symmetry) when $S_z$ is odd. The black symbols stand for the lowest energy states in each momentum sector or the first excited state of the cMR momentum sectors. We clearly observe the important dispersion of the (red) low energy states.}
\label{U0_2_U1_035}
\end{figure}

As mentioned in Sec.~\ref{sec:coupledMR}, the cMR has a ground state degeneracy that depends on the parity of $N/2$ and the $S_z$ sector. Focusing on the even values of $N$, there is an alternation of a three-fold degenerate ground state and a non-degenerate ground state depending on the parity of $S_z$. Away from the model interaction, we expect this degeneracy to be lifted while preserving this alternation and the correct momenta. This is indeed what is observed, the degeneracy being split into two different ways: within each $S_z$ sector and between the different $S_z$ sectors. This latest statement means that the low-energy manifold acquires a dispersion relation with respect to the spin projection along $z$. We show in Fig.~\ref{U0_2_U1_035} the energy spectrum for the largest system size that we have been able to reach (namely $N=14$) and in table the corresponding overlaps with the exact cMR states in each $S_z$ sectors. We have also checked that the PES provides the same phase identification (up to minor size effects) for the various system sizes. Note that due to the small number of system sizes that can be evaluated and to the parity effect over $N/2$, we cannot make any reliable extrapolation of the gap. In particular, it is not possible to see if the phase will become gapless.

\begin{table}[htbp]
\centering
\begin{tabular}{ l |c|c|c }
\hline
$S_z$ & $(0,0)$ & $(\pi,0)$ or $(0,\pi)$ & $(\pi,\pi)$\\
\hline
\hline
0 & $\text{--}$ & $\text{--}$ & $0.891$ \\
\hline
1 & $0.877$ & $0.885$ & $\text{--}$ \\
\hline
2 & $\text{--}$ & $\text{--}$ & $0.898$ \\
\hline
3 & $0.882$ & $0.887$ & $\text{--}$ \\
\hline
4 & $\text{--}$ & $\text{--}$ & $0.889$ \\
\hline
5 & $0.820$ & $0.875$ & $\text{--}$ \\
\hline
6 & $\text{--}$ & $\text{--}$ & $0.880$ \\
\hline
7 & $0.781$ & $0.832$ & $\text{--}$ \\
\hline
\end{tabular}
\caption{Overlaps for $N=14$ bosons between the cMR and the ground state of the Eq.~(\ref{hamil}) in the different momentum and $S_z$ sectors for $U_0=2.0$ and $U_1=0.35$. A dash symbol is used when the momentum sector is not compatible with a given $S_z$.}\label{tableoverlap}
\end{table}

As can be observed in Fig.~\ref{U0_2_U1_035}, the dispersion with $S_z$ can be rather important. If we want to be closer to the model cMR case, it might be desirable for this dispersion to be as flat as possible. Such a situation occurs in particular near the transition between the unpolarized regime and the fully unpolarized regime (see Fig.~\ref{Pz12}) which is also a region of high overlaps [see Figs.~\ref{cMR12}(a) and~\ref{cMR12}(b)]. A lower value of $U_1$ while keeping $U_0=2.0$ roughly offers such a situation. We can wonder if adding the Josephson coupling term of Eq.~(\ref{josephson}) would also drive the system into the $\pi/2$ rotated Halperin $(220)$ state discussed in Sec.~\ref{sec:coupledMR}. We have checked that this is indeed the case with a major difference : in our model, it requires a much larger value of $t_{\rm J}$ (typically around $t_{\rm J} \simeq 1$) compared to the cMR model Hamiltonian of Eq.~(\ref{hamiltonian32}). We can actually make a stronger statement: as shown in Ref.~\onlinecite{Moller-PhysRevB.90.235101}, with the proper amount of inter-layer interaction ($U_0=2.0$) and Josephson coupling ($t_{\rm J}=1$), and setting $U_1=0$, the resulting Hamiltonian is nothing but the model Hamiltonian for the $\pi/2$ rotated Halperin $(220)$ state. Adding some small $U_1$ or changing a little bit either $U_0$ or $t_{\rm J}$ does not change this picture.

\begin{figure*}
\centerline{\includegraphics[width=\linewidth]{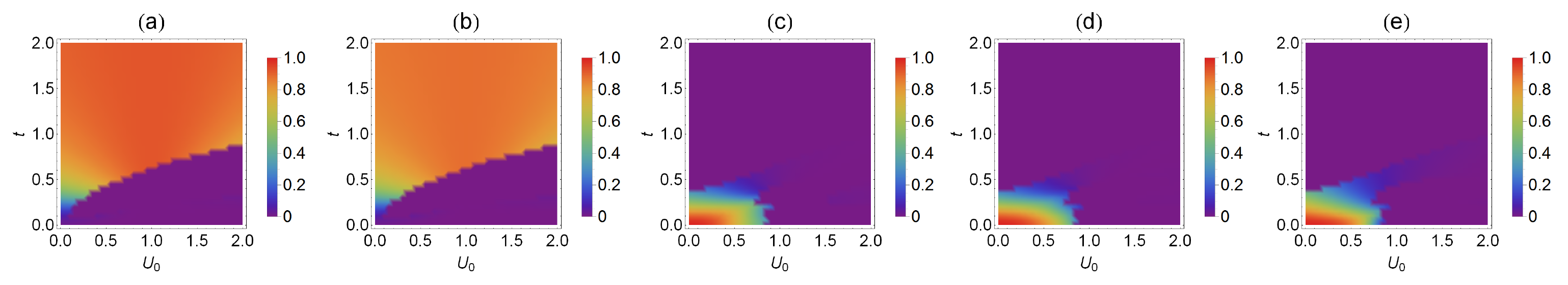}}
\caption{(Color online) Nature of the low energy manifold as a function of $U_0$ and $t$ for a system of $N=10$ bosons with $U_1=0$. The ground-state overlap with the exact MR state in (a) ${\bf k}=(0,0)$ and (b) ${\bf k}=(\pi,0)$ sector. (c)-(e) The ground-state overlap with the exact $(220)$ state in (c) ${\bf k}=(0,0)$, (d) ${\bf k}=(\pi,0)$ and (e) ${\bf k}=(\pi,\pi)$ sector.}
\label{overlapv0t}
\end{figure*}

\begin{figure}
\centerline{\includegraphics[width=\linewidth]{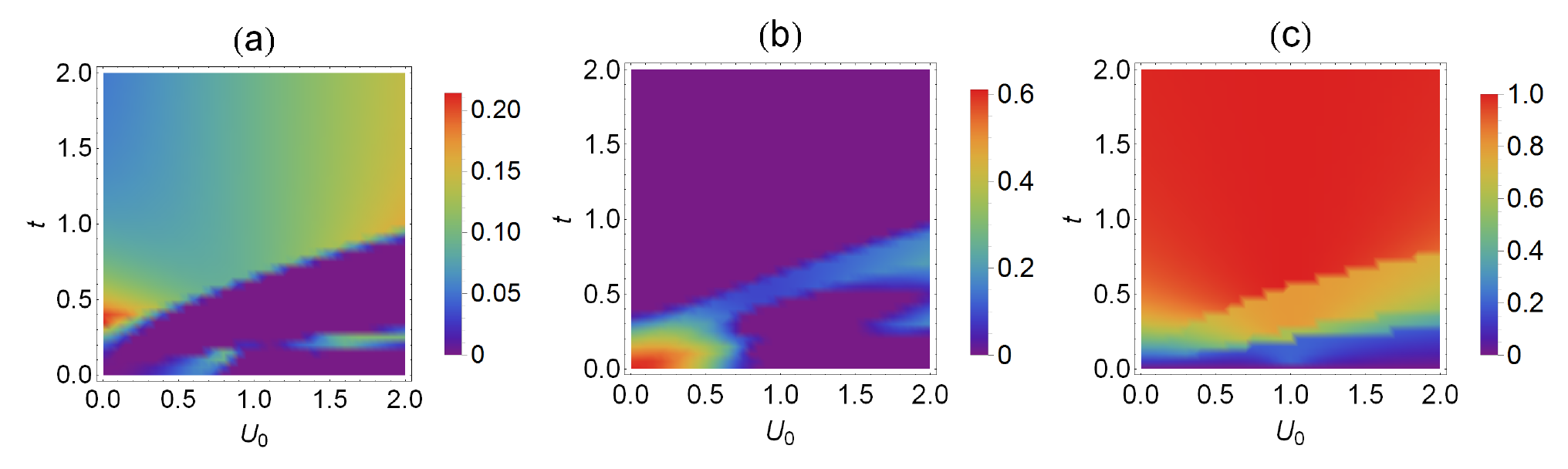}}
\caption{(Color online) Properties of the low-energy manifold as a function of $U_0$ and $t$ for a system of $N=10$ bosons with $U_1=0$. (a) The energy gap relative to the MR manifold. (b) The energy gap relative to the $(220)$ manifold. (c) The $x$ polarization ${\cal P}_x$ of the absolute ground state.}
\label{np_10}
\end{figure}

\begin{figure}
\centerline{\includegraphics[width=\linewidth]{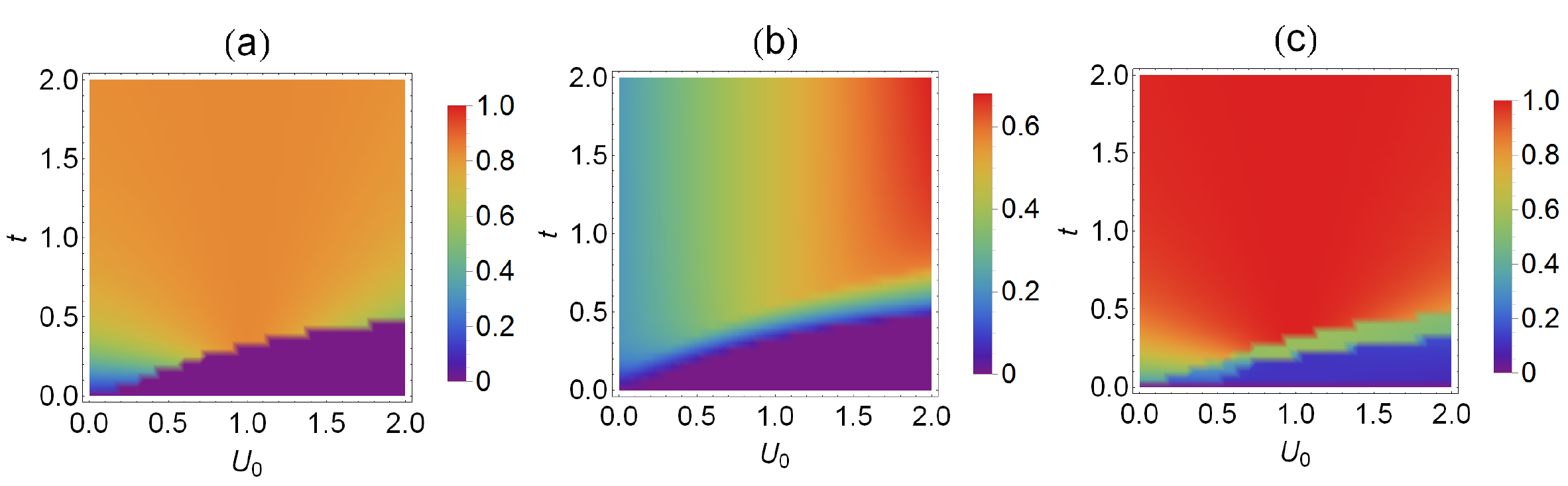}}
\caption{(Color online) Properties of the low-energy manifold as a function of $U_0$ and $t$ for a system of $N=9$ bosons with $U_1=0$. (a) The ground-state overlap with the exact MR state in the ${\bf k}=(0,0)$ sector. (b) The energy gap relative to the MR state. (c) The $x$ polarization ${\cal P}_x$ of the absolute ground state.}
\label{np_9}
\end{figure}

\section{Tunneling effect}\label{sec:phasediagramtunneling}

We now consider the effect of tunneling between the two layers. As mentioned in Sec.~\ref{sec:model}, the tunneling term acts as a Zeeman field along the fictitious $x$ axis. We thus expect that for a critical value of $t$, the system is polarized in this direction leading to an effective single-layer picture for the low-energy physics. As a consequence, the MR phase emerges beyond that critical tunneling amplitude. Beyond the determination of this critical value of $t$, we can wonder how it will be modified by the presence of the interlayer interaction. Here we only consider the role of $U_0$ and we set $U_1=0$. We already know from Sec.~\ref{sec:u0u1inter} that in the absence of tunneling, a moderate amount of $U_1$ drives the system into the MR phase whereas this phase does not appear along the $U_1=0$ line. Our choice allows to have more readable figures while capturing the relevant situations.

Due to the (generic) absence of any $SU(2)$ quantum number, the system sizes that can be simulated are smaller than those in Sec.~\ref{sec:phasediagraminter} (here up to $N=10$). At large tunneling, each orbital of the effective single-layer is made of the anti-symmetric (or symmetric depending on the sign of $t$) combination of an up layer and a down layer orbital. Away from this limit, the low-energy physics might still be described by a single-layer picture with a polarization axis that is still in the $x-z$ plane but that does not have to be along $x$ (see Ref.~\onlinecite{Thiebaut-PhysRevB.89.195421} for a detailed discussion). As a consequence, checking the nature of the low-energy states by computing overlaps with a single-layer model state (here the MR state) requires to rotate this latest in the $x-z$ plane and to find the angle $\theta\in[0,\pi/2]$ between the $z$ axis and the polarization axis that maximizes the overlap. This can be achieved by applying the spin rotation operator $R_y(\theta)=\prod_{j=1}^N e^{i \frac{\theta}{2}\sigma_y^j}$ which is the product of single-particle spin rotation operators, on the usual single-layer MR states to get the model MR states $|\Psi_{\textrm{MR}}(\theta)\rangle$ with rotation angle $\theta$. Then we search the optimal value of $\theta$ that maximizes the total overlap between the model MR state and the ground state of our system over momentum sectors $(0,0)$, $(\pi,0)$, and $(0,\pi)$ for even number of particles $N$.

We show in Figs.~\ref{overlapv0t}(a) and (b) the maximal overlap that can be reached for $N=10$ with the optimal rotation angle, which is almost always $\pi/2$ except at a few $(U_0,t)$ points. There is a large region of high overlap with the model MR state. Interestingly the inter-layer $U_0$ interaction enhances the overlap around the $U_0=1$ line by requiring a lower strength of $t$ to obtain a high overlap: $t\approx0.6$ is enough to reach overlap $\simeq0.9$ for $U_0=1$, while the overlap is still less at $t=2$ for $U_0=0$. Along the $U_0=1$ line, the interaction recovers its $SU(2)$ symmetry that is only partially broken by the tunneling term. In order to make sure the ground states (irrespective of their momenta) are in the MR momentum sectors, we also compute the energy gap (with the same definition as in Sec.~\ref{sec:u0u1inter}) relative to the MR manifold [Fig.~\ref{np_10}(a)]. The region with non-zero energy gap is consistent with that of high overlap. We can also look at the case where we have an odd number of particles. In Fig.~\ref{np_9}(a), we show the maximal overlap in the $(0,0)$ momentum sector between the ground state and the exact MR state obtained by spin rotation. The results are similar to those for an even particle number -- a similar large region of high overlap exists and the inter-layer interaction can further enhance the MR phase.

We have performed a similar study for the Halperin $(220)$ state. The overlaps with respect to this model state are given in Figs.~\ref{overlapv0t}(c), (d) and (e). The optimal rotation angle to reach the maximal overlap is almost $0$ everywhere. Compared with Figs.~\ref{overlapv0v1}(c), (d) (e), the region with high overlap shrinks in the $t$ direction, suggesting that the $(220)$ phase is more fragile under inter-layer tunneling $t$ than under the inter-layer $\mathcal{V}_1$ interaction. We have also checked the energy gap relative to the $(220)$ phase [Fig.~\ref{np_10}(b)], which gives consistent result with the overlap calculations. In a similar line of thought, we have looked at a possible realization of the cMR phase. Unfortunately, we did not find any strong signature of the cMR with a maximum overlap of $\approx 0.5$ around $U_0\approx 2.0$ and $t \approx 0.1$.

It is also instructive to compute the polarization along the $x$ axis for this system. For that purpose, we compute the $x$ polarization ${\cal P}_x= \langle \hat{S_x} \rangle / (N/2)$  of the system's absolute ground state (without focusing on specific quantum number sector) [Fig.~\ref{np_10}(c)]. As opposed to the polarization ${\cal P}_z$ studied in the Sec.~\ref{sec:u0u1inter}, the ground state is not an eigenstate of $\hat{S_x}$. The results show that once the system enters the MR phase, the $x$ polarization is close to $1$ as expected.

\section{Conclusion}
In this paper, we explore the phase diagram of a bilayer bosonic FQH system at total filling factor $\nu=1$ with Haldane's zero-order $U_0$ and first-order $U_1$ pseudopotential interactions and inter-layer tunneling. In the absence of tunneling, we have found strong signature of three phases: the Halperin $(220)$ state, the coupled Moore-Read state and a Moore-Read state. When the system is $SU(2)$ invariant, the MR phase becomes ferromagnetic for an inter-layer $U_1$ interaction larger than $U^c_1\approx0.2-0.3$, thus guaranteeing the MR phase in all $S_z$ sectors. Away from the $SU(2)$ invariant line $U_0=1$, there is still a very wide region of robust MR phase in the $S_z=0$ sector, and the system can fully polarize to a single-layer MR phase for $U_0>1$ and a inter-layer $U_1$ interaction larger than $U^c_1$. The presence of a region where the low-energy physics corresponds to the coupled Moore-Read state, offers a simpler realization of this phase involving only two-body interactions. The inter-layer tunneling drives the bosons into a single-component system polarized in the fictitious spin $x$ direction. For large enough tunneling, we observe the MR phase while the inter-layer interaction can further enhance these phases.

\begin{acknowledgments}
We thank D.~N.~Sheng, Wei Zhu and C.~von~Keyserlingk for discussions. N.~R. would like to thank T.~Neupert and G.~M\"oller for enlightening discussions. Z.~L. was supported by the Department of Energy, Office of Basic Energy Sciences through Grant No.~DE-SC0002140, and the Dahlem Research School (DRS) Postdoc Fellowship. N.~R. was supported by ANR-12-BS04-0002-02 and the Princeton Global Scholarship.
\end{acknowledgments}

\bibliography{bilayerboson}

\begin{thebibliography}{45}%
\makeatletter
\providecommand \@ifxundefined [1]{%
 \@ifx{#1\undefined}
}%
\providecommand \@ifnum [1]{%
 \ifnum #1\expandafter \@firstoftwo
 \else \expandafter \@secondoftwo
 \fi
}%
\providecommand \@ifx [1]{%
 \ifx #1\expandafter \@firstoftwo
 \else \expandafter \@secondoftwo
 \fi
}%
\providecommand \natexlab [1]{#1}%
\providecommand \enquote  [1]{``#1''}%
\providecommand \bibnamefont  [1]{#1}%
\providecommand \bibfnamefont [1]{#1}%
\providecommand \citenamefont [1]{#1}%
\providecommand \href@noop [0]{\@secondoftwo}%
\providecommand \href [0]{\begingroup \@sanitize@url \@href}%
\providecommand \@href[1]{\@@startlink{#1}\@@href}%
\providecommand \@@href[1]{\endgroup#1\@@endlink}%
\providecommand \@sanitize@url [0]{\catcode `\\12\catcode `\$12\catcode
  `\&12\catcode `\#12\catcode `\^12\catcode `\_12\catcode `\%12\relax}%
\providecommand \@@startlink[1]{}%
\providecommand \@@endlink[0]{}%
\providecommand \url  [0]{\begingroup\@sanitize@url \@url }%
\providecommand \@url [1]{\endgroup\@href {#1}{\urlprefix }}%
\providecommand \urlprefix  [0]{URL }%
\providecommand \Eprint [0]{\href }%
\providecommand \doibase [0]{http://dx.doi.org/}%
\providecommand \selectlanguage [0]{\@gobble}%
\providecommand \bibinfo  [0]{\@secondoftwo}%
\providecommand \bibfield  [0]{\@secondoftwo}%
\providecommand \translation [1]{[#1]}%
\providecommand \BibitemOpen [0]{}%
\providecommand \bibitemStop [0]{}%
\providecommand \bibitemNoStop [0]{.\EOS\space}%
\providecommand \EOS [0]{\spacefactor3000\relax}%
\providecommand \BibitemShut  [1]{\csname bibitem#1\endcsname}%
\let\auto@bib@innerbib\@empty
\bibitem [{\citenamefont {Bais}\ and\ \citenamefont
  {Slingerland}(2009)}]{Bais-PhysRevB.79.045316}%
  \BibitemOpen
  \bibfield  {author} {\bibinfo {author} {\bibfnamefont {F.~A.}\ \bibnamefont
  {Bais}}\ and\ \bibinfo {author} {\bibfnamefont {J.~K.}\ \bibnamefont
  {Slingerland}},\ }\href {\doibase 10.1103/PhysRevB.79.045316} {\bibfield
  {journal} {\bibinfo  {journal} {Phys. Rev. B}\ }\textbf {\bibinfo {volume}
  {79}},\ \bibinfo {pages} {045316} (\bibinfo {year} {2009})}\BibitemShut
  {NoStop}%
\bibitem [{\citenamefont {Barkeshli}\ and\ \citenamefont
  {Wen}(2010{\natexlab{a}})}]{Barkeshli-PhysRevLett.105.216804}%
  \BibitemOpen
  \bibfield  {author} {\bibinfo {author} {\bibfnamefont {M.}~\bibnamefont
  {Barkeshli}}\ and\ \bibinfo {author} {\bibfnamefont {X.-G.}\ \bibnamefont
  {Wen}},\ }\href {\doibase 10.1103/PhysRevLett.105.216804} {\bibfield
  {journal} {\bibinfo  {journal} {Phys. Rev. Lett.}\ }\textbf {\bibinfo
  {volume} {105}},\ \bibinfo {pages} {216804} (\bibinfo {year}
  {2010}{\natexlab{a}})}\BibitemShut {NoStop}%
\bibitem [{\citenamefont {{Cappelli}}\ \emph {et~al.}(1999)\citenamefont
  {{Cappelli}}, \citenamefont {{Georgiev}},\ and\ \citenamefont
  {{Todorov}}}]{Cappelli-1999CMaPh.205..657C}%
  \BibitemOpen
  \bibfield  {author} {\bibinfo {author} {\bibfnamefont {A.}~\bibnamefont
  {{Cappelli}}}, \bibinfo {author} {\bibfnamefont {L.~S.}\ \bibnamefont
  {{Georgiev}}}, \ and\ \bibinfo {author} {\bibfnamefont {I.~T.}\ \bibnamefont
  {{Todorov}}},\ }\href {\doibase 10.1007/s002200050693} {\bibfield  {journal}
  {\bibinfo  {journal} {Communications in Mathematical Physics}\ }\textbf
  {\bibinfo {volume} {205}},\ \bibinfo {pages} {657} (\bibinfo {year}
  {1999})}\BibitemShut {NoStop}%
\bibitem [{\citenamefont {{Froehlich}}\ \emph {et~al.}(2000)\citenamefont
  {{Froehlich}}, \citenamefont {{Pedrini}}, \citenamefont {{Schweigert}},\ and\
  \citenamefont {{Walcher}}}]{Froehlich-2000cond.mat..2330F}%
  \BibitemOpen
  \bibfield  {author} {\bibinfo {author} {\bibfnamefont {J.}~\bibnamefont
  {{Froehlich}}}, \bibinfo {author} {\bibfnamefont {B.}~\bibnamefont
  {{Pedrini}}}, \bibinfo {author} {\bibfnamefont {C.}~\bibnamefont
  {{Schweigert}}}, \ and\ \bibinfo {author} {\bibfnamefont {J.}~\bibnamefont
  {{Walcher}}},\ }\href@noop {} {\bibfield  {journal} {\bibinfo  {journal}
  {eprint arXiv:cond-mat/0002330}\ } (\bibinfo {year} {2000})},\ \Eprint
  {http://arxiv.org/abs/cond-mat/0002330} {cond-mat/0002330} \BibitemShut
  {NoStop}%
\bibitem [{\citenamefont {{Cappelli}}\ \emph {et~al.}(2001)\citenamefont
  {{Cappelli}}, \citenamefont {{Georgiev}},\ and\ \citenamefont
  {{Todorov}}}]{cappelli-2001NuPhB.599..499C}%
  \BibitemOpen
  \bibfield  {author} {\bibinfo {author} {\bibfnamefont {A.}~\bibnamefont
  {{Cappelli}}}, \bibinfo {author} {\bibfnamefont {L.~S.}\ \bibnamefont
  {{Georgiev}}}, \ and\ \bibinfo {author} {\bibfnamefont {I.~T.}\ \bibnamefont
  {{Todorov}}},\ }\href {\doibase 10.1016/S0550-3213(00)00774-4} {\bibfield
  {journal} {\bibinfo  {journal} {Nuclear Physics B}\ }\textbf {\bibinfo
  {volume} {599}},\ \bibinfo {pages} {499} (\bibinfo {year}
  {2001})}\BibitemShut {NoStop}%
\bibitem [{\citenamefont {Barkeshli}\ and\ \citenamefont
  {Wen}(2010{\natexlab{b}})}]{Barkeshli-PhysRevB.81.155302}%
  \BibitemOpen
  \bibfield  {author} {\bibinfo {author} {\bibfnamefont {M.}~\bibnamefont
  {Barkeshli}}\ and\ \bibinfo {author} {\bibfnamefont {X.-G.}\ \bibnamefont
  {Wen}},\ }\href {\doibase 10.1103/PhysRevB.81.155302} {\bibfield  {journal}
  {\bibinfo  {journal} {Phys. Rev. B}\ }\textbf {\bibinfo {volume} {81}},\
  \bibinfo {pages} {155302} (\bibinfo {year} {2010}{\natexlab{b}})}\BibitemShut
  {NoStop}%
\bibitem [{\citenamefont {Repellin}\ \emph {et~al.}(2015)\citenamefont
  {Repellin}, \citenamefont {Neupert}, \citenamefont {Bernevig},\ and\
  \citenamefont {Regnault}}]{Repellin-PhysRevB.92.115128}%
  \BibitemOpen
  \bibfield  {author} {\bibinfo {author} {\bibfnamefont {C.}~\bibnamefont
  {Repellin}}, \bibinfo {author} {\bibfnamefont {T.}~\bibnamefont {Neupert}},
  \bibinfo {author} {\bibfnamefont {B.~A.}\ \bibnamefont {Bernevig}}, \ and\
  \bibinfo {author} {\bibfnamefont {N.}~\bibnamefont {Regnault}},\ }\href
  {\doibase 10.1103/PhysRevB.92.115128} {\bibfield  {journal} {\bibinfo
  {journal} {Phys. Rev. B}\ }\textbf {\bibinfo {volume} {92}},\ \bibinfo
  {pages} {115128} (\bibinfo {year} {2015})}\BibitemShut {NoStop}%
\bibitem [{\citenamefont {Laughlin}(1983)}]{Laughlin-PhysRevLett.50.1395}%
  \BibitemOpen
  \bibfield  {author} {\bibinfo {author} {\bibfnamefont {R.~B.}\ \bibnamefont
  {Laughlin}},\ }\href {\doibase 10.1103/PhysRevLett.50.1395} {\bibfield
  {journal} {\bibinfo  {journal} {Phys. Rev. Lett.}\ }\textbf {\bibinfo
  {volume} {50}},\ \bibinfo {pages} {1395} (\bibinfo {year}
  {1983})}\BibitemShut {NoStop}%
\bibitem [{\citenamefont {Moore}\ and\ \citenamefont
  {Read}(1991)}]{Moore1991362}%
  \BibitemOpen
  \bibfield  {author} {\bibinfo {author} {\bibfnamefont {G.}~\bibnamefont
  {Moore}}\ and\ \bibinfo {author} {\bibfnamefont {N.}~\bibnamefont {Read}},\
  }\href {\doibase http://dx.doi.org/10.1016/0550-3213(91)90407-O} {\bibfield
  {journal} {\bibinfo  {journal} {Nuclear Physics B}\ }\textbf {\bibinfo
  {volume} {360}},\ \bibinfo {pages} {362 } (\bibinfo {year}
  {1991})}\BibitemShut {NoStop}%
\bibitem [{\citenamefont {{Fradkin}}\ \emph {et~al.}(1998)\citenamefont
  {{Fradkin}}, \citenamefont {{Nayak}}, \citenamefont {{Tsvelik}},\ and\
  \citenamefont {{Wilczek}}}]{Fradkin-1998NuPhB}%
  \BibitemOpen
  \bibfield  {author} {\bibinfo {author} {\bibfnamefont {E.}~\bibnamefont
  {{Fradkin}}}, \bibinfo {author} {\bibfnamefont {C.}~\bibnamefont {{Nayak}}},
  \bibinfo {author} {\bibfnamefont {A.}~\bibnamefont {{Tsvelik}}}, \ and\
  \bibinfo {author} {\bibfnamefont {F.}~\bibnamefont {{Wilczek}}},\ }\href
  {\doibase 10.1016/S0550-3213(98)00111-4} {\bibfield  {journal} {\bibinfo
  {journal} {Nuclear Physics B}\ }\textbf {\bibinfo {volume} {516}},\ \bibinfo
  {pages} {704} (\bibinfo {year} {1998})},\ \Eprint
  {http://arxiv.org/abs/cond-mat/9711087} {cond-mat/9711087} \BibitemShut
  {NoStop}%
\bibitem [{\citenamefont {{Fradkin}}\ \emph {et~al.}(1999)\citenamefont
  {{Fradkin}}, \citenamefont {{Nayak}},\ and\ \citenamefont
  {{Schoutens}}}]{Fradkin-1999NuPhB}%
  \BibitemOpen
  \bibfield  {author} {\bibinfo {author} {\bibfnamefont {E.}~\bibnamefont
  {{Fradkin}}}, \bibinfo {author} {\bibfnamefont {C.}~\bibnamefont {{Nayak}}},
  \ and\ \bibinfo {author} {\bibfnamefont {K.}~\bibnamefont {{Schoutens}}},\
  }\href {\doibase 10.1016/S0550-3213(99)00039-5} {\bibfield  {journal}
  {\bibinfo  {journal} {Nuclear Physics B}\ }\textbf {\bibinfo {volume}
  {546}},\ \bibinfo {pages} {711} (\bibinfo {year} {1999})},\ \Eprint
  {http://arxiv.org/abs/cond-mat/9811005} {cond-mat/9811005} \BibitemShut
  {NoStop}%
\bibitem [{\citenamefont {Read}\ and\ \citenamefont
  {Green}(2000)}]{read-green-PhysRevB.61.10267}%
  \BibitemOpen
  \bibfield  {author} {\bibinfo {author} {\bibfnamefont {N.}~\bibnamefont
  {Read}}\ and\ \bibinfo {author} {\bibfnamefont {D.}~\bibnamefont {Green}},\
  }\href {\doibase 10.1103/PhysRevB.61.10267} {\bibfield  {journal} {\bibinfo
  {journal} {Phys. Rev. B}\ }\textbf {\bibinfo {volume} {61}},\ \bibinfo
  {pages} {10267} (\bibinfo {year} {2000})}\BibitemShut {NoStop}%
\bibitem [{\citenamefont {Cooper}\ \emph {et~al.}(2001)\citenamefont {Cooper},
  \citenamefont {Wilkin},\ and\ \citenamefont
  {Gunn}}]{Cooper-PhysRevLett.87.120405}%
  \BibitemOpen
  \bibfield  {author} {\bibinfo {author} {\bibfnamefont {N.~R.}\ \bibnamefont
  {Cooper}}, \bibinfo {author} {\bibfnamefont {N.~K.}\ \bibnamefont {Wilkin}},
  \ and\ \bibinfo {author} {\bibfnamefont {J.~M.~F.}\ \bibnamefont {Gunn}},\
  }\href {\doibase 10.1103/PhysRevLett.87.120405} {\bibfield  {journal}
  {\bibinfo  {journal} {Phys. Rev. Lett.}\ }\textbf {\bibinfo {volume} {87}},\
  \bibinfo {pages} {120405} (\bibinfo {year} {2001})}\BibitemShut {NoStop}%
\bibitem [{\citenamefont {Regnault}\ and\ \citenamefont
  {Jolicoeur}(2003)}]{regnault-PhysRevLett.91.030402}%
  \BibitemOpen
  \bibfield  {author} {\bibinfo {author} {\bibfnamefont {N.}~\bibnamefont
  {Regnault}}\ and\ \bibinfo {author} {\bibfnamefont {T.}~\bibnamefont
  {Jolicoeur}},\ }\href {\doibase 10.1103/PhysRevLett.91.030402} {\bibfield
  {journal} {\bibinfo  {journal} {Phys. Rev. Lett.}\ }\textbf {\bibinfo
  {volume} {91}},\ \bibinfo {pages} {030402} (\bibinfo {year}
  {2003})}\BibitemShut {NoStop}%
\bibitem [{\citenamefont {Chang}\ \emph {et~al.}(2005)\citenamefont {Chang},
  \citenamefont {Regnault}, \citenamefont {Jolicoeur},\ and\ \citenamefont
  {Jain}}]{Chang-PhysRevA.72.013611}%
  \BibitemOpen
  \bibfield  {author} {\bibinfo {author} {\bibfnamefont {C.-C.}\ \bibnamefont
  {Chang}}, \bibinfo {author} {\bibfnamefont {N.}~\bibnamefont {Regnault}},
  \bibinfo {author} {\bibfnamefont {T.}~\bibnamefont {Jolicoeur}}, \ and\
  \bibinfo {author} {\bibfnamefont {J.~K.}\ \bibnamefont {Jain}},\ }\href
  {\doibase 10.1103/PhysRevA.72.013611} {\bibfield  {journal} {\bibinfo
  {journal} {Phys. Rev. A}\ }\textbf {\bibinfo {volume} {72}},\ \bibinfo
  {pages} {013611} (\bibinfo {year} {2005})}\BibitemShut {NoStop}%
\bibitem [{\citenamefont {Regnault}\ and\ \citenamefont
  {Jolicoeur}(2004)}]{PhysRevB.69.235309}%
  \BibitemOpen
  \bibfield  {author} {\bibinfo {author} {\bibfnamefont {N.}~\bibnamefont
  {Regnault}}\ and\ \bibinfo {author} {\bibfnamefont {T.}~\bibnamefont
  {Jolicoeur}},\ }\href {\doibase 10.1103/PhysRevB.69.235309} {\bibfield
  {journal} {\bibinfo  {journal} {Phys. Rev. B}\ }\textbf {\bibinfo {volume}
  {69}},\ \bibinfo {pages} {235309} (\bibinfo {year} {2004})}\BibitemShut
  {NoStop}%
\bibitem [{\citenamefont {Regnault}\ and\ \citenamefont
  {Jolicoeur}(2007)}]{PhysRevB.76.235324}%
  \BibitemOpen
  \bibfield  {author} {\bibinfo {author} {\bibfnamefont {N.}~\bibnamefont
  {Regnault}}\ and\ \bibinfo {author} {\bibfnamefont {T.}~\bibnamefont
  {Jolicoeur}},\ }\href {\doibase 10.1103/PhysRevB.76.235324} {\bibfield
  {journal} {\bibinfo  {journal} {Phys. Rev. B}\ }\textbf {\bibinfo {volume}
  {76}},\ \bibinfo {pages} {235324} (\bibinfo {year} {2007})}\BibitemShut
  {NoStop}%
\bibitem [{\citenamefont {Wu}\ and\ \citenamefont
  {Jain}(2013)}]{Wu-PhysRevB.87.245123}%
  \BibitemOpen
  \bibfield  {author} {\bibinfo {author} {\bibfnamefont {Y.-H.}\ \bibnamefont
  {Wu}}\ and\ \bibinfo {author} {\bibfnamefont {J.~K.}\ \bibnamefont {Jain}},\
  }\href {\doibase 10.1103/PhysRevB.87.245123} {\bibfield  {journal} {\bibinfo
  {journal} {Phys. Rev. B}\ }\textbf {\bibinfo {volume} {87}},\ \bibinfo
  {pages} {245123} (\bibinfo {year} {2013})}\BibitemShut {NoStop}%
\bibitem [{\citenamefont {M\"oller}\ \emph {et~al.}(2014)\citenamefont
  {M\"oller}, \citenamefont {Hormozi}, \citenamefont {Slingerland},\ and\
  \citenamefont {Simon}}]{Moller-PhysRevB.90.235101}%
  \BibitemOpen
  \bibfield  {author} {\bibinfo {author} {\bibfnamefont {G.}~\bibnamefont
  {M\"oller}}, \bibinfo {author} {\bibfnamefont {L.}~\bibnamefont {Hormozi}},
  \bibinfo {author} {\bibfnamefont {J.}~\bibnamefont {Slingerland}}, \ and\
  \bibinfo {author} {\bibfnamefont {S.~H.}\ \bibnamefont {Simon}},\ }\href
  {\doibase 10.1103/PhysRevB.90.235101} {\bibfield  {journal} {\bibinfo
  {journal} {Phys. Rev. B}\ }\textbf {\bibinfo {volume} {90}},\ \bibinfo
  {pages} {235101} (\bibinfo {year} {2014})}\BibitemShut {NoStop}%
\bibitem [{\citenamefont {Wu}\ and\ \citenamefont
  {Jain}(2015)}]{PhysRevA.91.063623}%
  \BibitemOpen
  \bibfield  {author} {\bibinfo {author} {\bibfnamefont {Y.-H.}\ \bibnamefont
  {Wu}}\ and\ \bibinfo {author} {\bibfnamefont {J.~K.}\ \bibnamefont {Jain}},\
  }\href {\doibase 10.1103/PhysRevA.91.063623} {\bibfield  {journal} {\bibinfo
  {journal} {Phys. Rev. A}\ }\textbf {\bibinfo {volume} {91}},\ \bibinfo
  {pages} {063623} (\bibinfo {year} {2015})}\BibitemShut {NoStop}%
\bibitem [{\citenamefont {Repellin}\ \emph {et~al.}(2014)\citenamefont
  {Repellin}, \citenamefont {Bernevig},\ and\ \citenamefont
  {Regnault}}]{PhysRevB.90.245401}%
  \BibitemOpen
  \bibfield  {author} {\bibinfo {author} {\bibfnamefont {C.}~\bibnamefont
  {Repellin}}, \bibinfo {author} {\bibfnamefont {B.~A.}\ \bibnamefont
  {Bernevig}}, \ and\ \bibinfo {author} {\bibfnamefont {N.}~\bibnamefont
  {Regnault}},\ }\href {\doibase 10.1103/PhysRevB.90.245401} {\bibfield
  {journal} {\bibinfo  {journal} {Phys. Rev. B}\ }\textbf {\bibinfo {volume}
  {90}},\ \bibinfo {pages} {245401} (\bibinfo {year} {2014})}\BibitemShut
  {NoStop}%
\bibitem [{\citenamefont {Zhu}\ \emph {et~al.}(2015)\citenamefont {Zhu},
  \citenamefont {Gong}, \citenamefont {Sheng},\ and\ \citenamefont
  {Sheng}}]{Zhu-PhysRevB.91.245126}%
  \BibitemOpen
  \bibfield  {author} {\bibinfo {author} {\bibfnamefont {W.}~\bibnamefont
  {Zhu}}, \bibinfo {author} {\bibfnamefont {S.~S.}\ \bibnamefont {Gong}},
  \bibinfo {author} {\bibfnamefont {D.~N.}\ \bibnamefont {Sheng}}, \ and\
  \bibinfo {author} {\bibfnamefont {L.}~\bibnamefont {Sheng}},\ }\href
  {\doibase 10.1103/PhysRevB.91.245126} {\bibfield  {journal} {\bibinfo
  {journal} {Phys. Rev. B}\ }\textbf {\bibinfo {volume} {91}},\ \bibinfo
  {pages} {245126} (\bibinfo {year} {2015})}\BibitemShut {NoStop}%
\bibitem [{\citenamefont {Ardonne}\ and\ \citenamefont
  {Schoutens}(1999)}]{Ardonne-PhysRevLett.82.5096}%
  \BibitemOpen
  \bibfield  {author} {\bibinfo {author} {\bibfnamefont {E.}~\bibnamefont
  {Ardonne}}\ and\ \bibinfo {author} {\bibfnamefont {K.}~\bibnamefont
  {Schoutens}},\ }\href {\doibase 10.1103/PhysRevLett.82.5096} {\bibfield
  {journal} {\bibinfo  {journal} {Phys. Rev. Lett.}\ }\textbf {\bibinfo
  {volume} {82}},\ \bibinfo {pages} {5096} (\bibinfo {year}
  {1999})}\BibitemShut {NoStop}%
\bibitem [{\citenamefont {Furukawa}\ and\ \citenamefont
  {Ueda}(2013)}]{Furukawa-PhysRevLett.111.090401}%
  \BibitemOpen
  \bibfield  {author} {\bibinfo {author} {\bibfnamefont {S.}~\bibnamefont
  {Furukawa}}\ and\ \bibinfo {author} {\bibfnamefont {M.}~\bibnamefont
  {Ueda}},\ }\href {\doibase 10.1103/PhysRevLett.111.090401} {\bibfield
  {journal} {\bibinfo  {journal} {Phys. Rev. Lett.}\ }\textbf {\bibinfo
  {volume} {111}},\ \bibinfo {pages} {090401} (\bibinfo {year}
  {2013})}\BibitemShut {NoStop}%
\bibitem [{\citenamefont {Regnault}\ and\ \citenamefont
  {Senthil}(2013)}]{Regnault-PhysRevB.88.161106}%
  \BibitemOpen
  \bibfield  {author} {\bibinfo {author} {\bibfnamefont {N.}~\bibnamefont
  {Regnault}}\ and\ \bibinfo {author} {\bibfnamefont {T.}~\bibnamefont
  {Senthil}},\ }\href {\doibase 10.1103/PhysRevB.88.161106} {\bibfield
  {journal} {\bibinfo  {journal} {Phys. Rev. B}\ }\textbf {\bibinfo {volume}
  {88}},\ \bibinfo {pages} {161106} (\bibinfo {year} {2013})}\BibitemShut
  {NoStop}%
\bibitem [{\citenamefont {Papi\'{c}}\ \emph {et~al.}(2010)\citenamefont
  {Papi\'{c}}, \citenamefont {Goerbig}, \citenamefont {Regnault},\ and\
  \citenamefont {Milovanovi\'{c}}}]{Papic-PhysRevB.82.075302}%
  \BibitemOpen
  \bibfield  {author} {\bibinfo {author} {\bibfnamefont {Z.}~\bibnamefont
  {Papi\'{c}}}, \bibinfo {author} {\bibfnamefont {M.~O.}\ \bibnamefont
  {Goerbig}}, \bibinfo {author} {\bibfnamefont {N.}~\bibnamefont {Regnault}}, \
  and\ \bibinfo {author} {\bibfnamefont {M.~V.}\ \bibnamefont
  {Milovanovi\'{c}}},\ }\href {\doibase 10.1103/PhysRevB.82.075302} {\bibfield
  {journal} {\bibinfo  {journal} {Phys. Rev. B}\ }\textbf {\bibinfo {volume}
  {82}},\ \bibinfo {pages} {075302} (\bibinfo {year} {2010})}\BibitemShut
  {NoStop}%
\bibitem [{\citenamefont {Halperin}(1983)}]{halperin1983theory}%
  \BibitemOpen
  \bibfield  {author} {\bibinfo {author} {\bibfnamefont {B.~I.}\ \bibnamefont
  {Halperin}},\ }\href {\doibase 10.5169/seals-115362} {\bibfield  {journal}
  {\bibinfo  {journal} {Helv. Phys. Acta}\ }\textbf {\bibinfo {volume} {56}},\
  \bibinfo {pages} {75} (\bibinfo {year} {1983})}\BibitemShut {NoStop}%
\bibitem [{\citenamefont {{Rezayi}}\ \emph {et~al.}(2010)\citenamefont
  {{Rezayi}}, \citenamefont {{Wen}},\ and\ \citenamefont
  {{Read}}}]{Rezayi-2010arXiv1007.2022R}%
  \BibitemOpen
  \bibfield  {author} {\bibinfo {author} {\bibfnamefont {E.}~\bibnamefont
  {{Rezayi}}}, \bibinfo {author} {\bibfnamefont {X.-G.}\ \bibnamefont {{Wen}}},
  \ and\ \bibinfo {author} {\bibfnamefont {N.}~\bibnamefont {{Read}}},\
  }\href@noop {} {\bibfield  {journal} {\bibinfo  {journal} {ArXiv e-prints}\ }
  (\bibinfo {year} {2010})},\ \Eprint {http://arxiv.org/abs/1007.2022}
  {arXiv:1007.2022 [cond-mat.mes-hall]} \BibitemShut {NoStop}%
\bibitem [{\citenamefont {Peterson}\ \emph {et~al.}(2015)\citenamefont
  {Peterson}, \citenamefont {Wu}, \citenamefont {Cheng}, \citenamefont
  {Barkeshli}, \citenamefont {Wang},\ and\ \citenamefont
  {Das~Sarma}}]{Peterson-PhysRevB.92.035103}%
  \BibitemOpen
  \bibfield  {author} {\bibinfo {author} {\bibfnamefont {M.~R.}\ \bibnamefont
  {Peterson}}, \bibinfo {author} {\bibfnamefont {Y.-L.}\ \bibnamefont {Wu}},
  \bibinfo {author} {\bibfnamefont {M.}~\bibnamefont {Cheng}}, \bibinfo
  {author} {\bibfnamefont {M.}~\bibnamefont {Barkeshli}}, \bibinfo {author}
  {\bibfnamefont {Z.}~\bibnamefont {Wang}}, \ and\ \bibinfo {author}
  {\bibfnamefont {S.}~\bibnamefont {Das~Sarma}},\ }\href {\doibase
  10.1103/PhysRevB.92.035103} {\bibfield  {journal} {\bibinfo  {journal} {Phys.
  Rev. B}\ }\textbf {\bibinfo {volume} {92}},\ \bibinfo {pages} {035103}
  (\bibinfo {year} {2015})}\BibitemShut {NoStop}%
\bibitem [{\citenamefont {Vaezi}\ and\ \citenamefont
  {Barkeshli}(2014)}]{Vaezi-PhysRevLett.113.236804}%
  \BibitemOpen
  \bibfield  {author} {\bibinfo {author} {\bibfnamefont {A.}~\bibnamefont
  {Vaezi}}\ and\ \bibinfo {author} {\bibfnamefont {M.}~\bibnamefont
  {Barkeshli}},\ }\href {\doibase 10.1103/PhysRevLett.113.236804} {\bibfield
  {journal} {\bibinfo  {journal} {Phys. Rev. Lett.}\ }\textbf {\bibinfo
  {volume} {113}},\ \bibinfo {pages} {236804} (\bibinfo {year}
  {2014})}\BibitemShut {NoStop}%
\bibitem [{\citenamefont {Liu}\ \emph {et~al.}(2015)\citenamefont {Liu},
  \citenamefont {Vaezi}, \citenamefont {Lee},\ and\ \citenamefont
  {Kim}}]{Liu-PhysRevB.92.081102}%
  \BibitemOpen
  \bibfield  {author} {\bibinfo {author} {\bibfnamefont {Z.}~\bibnamefont
  {Liu}}, \bibinfo {author} {\bibfnamefont {A.}~\bibnamefont {Vaezi}}, \bibinfo
  {author} {\bibfnamefont {K.}~\bibnamefont {Lee}}, \ and\ \bibinfo {author}
  {\bibfnamefont {E.-A.}\ \bibnamefont {Kim}},\ }\href {\doibase
  10.1103/PhysRevB.92.081102} {\bibfield  {journal} {\bibinfo  {journal} {Phys.
  Rev. B}\ }\textbf {\bibinfo {volume} {92}},\ \bibinfo {pages} {081102}
  (\bibinfo {year} {2015})}\BibitemShut {NoStop}%
\bibitem [{\citenamefont {Ardonne}\ \emph {et~al.}(2002)\citenamefont
  {Ardonne}, \citenamefont {Lankvelt}, \citenamefont {Ludwig},\ and\
  \citenamefont {Schoutens}}]{Ardonne-2002}%
  \BibitemOpen
  \bibfield  {author} {\bibinfo {author} {\bibfnamefont {E.}~\bibnamefont
  {Ardonne}}, \bibinfo {author} {\bibfnamefont {F.~J. M.~v.}\ \bibnamefont
  {Lankvelt}}, \bibinfo {author} {\bibfnamefont {A.~W.~W.}\ \bibnamefont
  {Ludwig}}, \ and\ \bibinfo {author} {\bibfnamefont {K.}~\bibnamefont
  {Schoutens}},\ }\href {\doibase 10.1103/PhysRevB.65.041305} {\bibfield
  {journal} {\bibinfo  {journal} {Phys. Rev. B}\ }\textbf {\bibinfo {volume}
  {65}},\ \bibinfo {pages} {041305} (\bibinfo {year} {2002})}\BibitemShut
  {NoStop}%
\bibitem [{\citenamefont {Geraedts}\ \emph {et~al.}(2015)\citenamefont
  {Geraedts}, \citenamefont {Zaletel}, \citenamefont
  {Papi\ifmmode~\acute{c}\else \'{c}\fi{}},\ and\ \citenamefont
  {Mong}}]{Geraedts-PhysRevB.91.205139}%
  \BibitemOpen
  \bibfield  {author} {\bibinfo {author} {\bibfnamefont {S.}~\bibnamefont
  {Geraedts}}, \bibinfo {author} {\bibfnamefont {M.~P.}\ \bibnamefont
  {Zaletel}}, \bibinfo {author} {\bibfnamefont {Z.}~\bibnamefont
  {Papi\ifmmode~\acute{c}\else \'{c}\fi{}}}, \ and\ \bibinfo {author}
  {\bibfnamefont {R.~S.~K.}\ \bibnamefont {Mong}},\ }\href {\doibase
  10.1103/PhysRevB.91.205139} {\bibfield  {journal} {\bibinfo  {journal} {Phys.
  Rev. B}\ }\textbf {\bibinfo {volume} {91}},\ \bibinfo {pages} {205139}
  (\bibinfo {year} {2015})}\BibitemShut {NoStop}%
\bibitem [{\citenamefont {Hormozi}\ \emph {et~al.}(2012)\citenamefont
  {Hormozi}, \citenamefont {M\"oller},\ and\ \citenamefont
  {Simon}}]{Hormozi-PhysRevLett.108.256809}%
  \BibitemOpen
  \bibfield  {author} {\bibinfo {author} {\bibfnamefont {L.}~\bibnamefont
  {Hormozi}}, \bibinfo {author} {\bibfnamefont {G.}~\bibnamefont {M\"oller}}, \
  and\ \bibinfo {author} {\bibfnamefont {S.~H.}\ \bibnamefont {Simon}},\ }\href
  {\doibase 10.1103/PhysRevLett.108.256809} {\bibfield  {journal} {\bibinfo
  {journal} {Phys. Rev. Lett.}\ }\textbf {\bibinfo {volume} {108}},\ \bibinfo
  {pages} {256809} (\bibinfo {year} {2012})}\BibitemShut {NoStop}%
\bibitem [{\citenamefont {Haldane}(1983)}]{Haldane-PhysRevLett.51.605}%
  \BibitemOpen
  \bibfield  {author} {\bibinfo {author} {\bibfnamefont {F.~D.~M.}\
  \bibnamefont {Haldane}},\ }\href {\doibase 10.1103/PhysRevLett.51.605}
  {\bibfield  {journal} {\bibinfo  {journal} {Phys. Rev. Lett.}\ }\textbf
  {\bibinfo {volume} {51}},\ \bibinfo {pages} {605} (\bibinfo {year}
  {1983})}\BibitemShut {NoStop}%
\bibitem [{\citenamefont {Haldane}(1985)}]{Haldane85-PhysRevLett.55.2095}%
  \BibitemOpen
  \bibfield  {author} {\bibinfo {author} {\bibfnamefont {F.~D.~M.}\
  \bibnamefont {Haldane}},\ }\href {\doibase 10.1103/PhysRevLett.55.2095}
  {\bibfield  {journal} {\bibinfo  {journal} {Phys. Rev. Lett.}\ }\textbf
  {\bibinfo {volume} {55}},\ \bibinfo {pages} {2095} (\bibinfo {year}
  {1985})}\BibitemShut {NoStop}%
\bibitem [{\citenamefont {Yang}\ \emph {et~al.}(2012)\citenamefont {Yang},
  \citenamefont {Hu}, \citenamefont {Papi\ifmmode~\acute{c}\else \'{c}\fi{}},\
  and\ \citenamefont {Haldane}}]{Yang-PhysRevLett.108.256807}%
  \BibitemOpen
  \bibfield  {author} {\bibinfo {author} {\bibfnamefont {B.}~\bibnamefont
  {Yang}}, \bibinfo {author} {\bibfnamefont {Z.-X.}\ \bibnamefont {Hu}},
  \bibinfo {author} {\bibfnamefont {Z.}~\bibnamefont
  {Papi\ifmmode~\acute{c}\else \'{c}\fi{}}}, \ and\ \bibinfo {author}
  {\bibfnamefont {F.~D.~M.}\ \bibnamefont {Haldane}},\ }\href {\doibase
  10.1103/PhysRevLett.108.256807} {\bibfield  {journal} {\bibinfo  {journal}
  {Phys. Rev. Lett.}\ }\textbf {\bibinfo {volume} {108}},\ \bibinfo {pages}
  {256807} (\bibinfo {year} {2012})}\BibitemShut {NoStop}%
\bibitem [{\citenamefont {M\"oller}\ \emph {et~al.}(2011)\citenamefont
  {M\"oller}, \citenamefont {W\'ojs},\ and\ \citenamefont
  {Cooper}}]{moller-PhysRevLett.107.036803}%
  \BibitemOpen
  \bibfield  {author} {\bibinfo {author} {\bibfnamefont {G.}~\bibnamefont
  {M\"oller}}, \bibinfo {author} {\bibfnamefont {A.}~\bibnamefont {W\'ojs}}, \
  and\ \bibinfo {author} {\bibfnamefont {N.~R.}\ \bibnamefont {Cooper}},\
  }\href {\doibase 10.1103/PhysRevLett.107.036803} {\bibfield  {journal}
  {\bibinfo  {journal} {Phys. Rev. Lett.}\ }\textbf {\bibinfo {volume} {107}},\
  \bibinfo {pages} {036803} (\bibinfo {year} {2011})}\BibitemShut {NoStop}%
\bibitem [{\citenamefont {Bonderson}\ \emph {et~al.}(2011)\citenamefont
  {Bonderson}, \citenamefont {Feiguin},\ and\ \citenamefont
  {Nayak}}]{bonderson-PhysRevLett.106.186802}%
  \BibitemOpen
  \bibfield  {author} {\bibinfo {author} {\bibfnamefont {P.}~\bibnamefont
  {Bonderson}}, \bibinfo {author} {\bibfnamefont {A.~E.}\ \bibnamefont
  {Feiguin}}, \ and\ \bibinfo {author} {\bibfnamefont {C.}~\bibnamefont
  {Nayak}},\ }\href {\doibase 10.1103/PhysRevLett.106.186802} {\bibfield
  {journal} {\bibinfo  {journal} {Phys. Rev. Lett.}\ }\textbf {\bibinfo
  {volume} {106}},\ \bibinfo {pages} {186802} (\bibinfo {year}
  {2011})}\BibitemShut {NoStop}%
\bibitem [{\citenamefont {Sterdyniak}\ \emph {et~al.}(2011)\citenamefont
  {Sterdyniak}, \citenamefont {Regnault},\ and\ \citenamefont
  {Bernevig}}]{PhysRevLett.106.100405}%
  \BibitemOpen
  \bibfield  {author} {\bibinfo {author} {\bibfnamefont {A.}~\bibnamefont
  {Sterdyniak}}, \bibinfo {author} {\bibfnamefont {N.}~\bibnamefont
  {Regnault}}, \ and\ \bibinfo {author} {\bibfnamefont {B.~A.}\ \bibnamefont
  {Bernevig}},\ }\href {\doibase 10.1103/PhysRevLett.106.100405} {\bibfield
  {journal} {\bibinfo  {journal} {Phys. Rev. Lett.}\ }\textbf {\bibinfo
  {volume} {106}},\ \bibinfo {pages} {100405} (\bibinfo {year}
  {2011})}\BibitemShut {NoStop}%
\bibitem [{\citenamefont {Bernevig}\ and\ \citenamefont
  {Haldane}(2008)}]{PhysRevLett.100.246802}%
  \BibitemOpen
  \bibfield  {author} {\bibinfo {author} {\bibfnamefont {B.~A.}\ \bibnamefont
  {Bernevig}}\ and\ \bibinfo {author} {\bibfnamefont {F.~D.~M.}\ \bibnamefont
  {Haldane}},\ }\href {\doibase 10.1103/PhysRevLett.100.246802} {\bibfield
  {journal} {\bibinfo  {journal} {Phys. Rev. Lett.}\ }\textbf {\bibinfo
  {volume} {100}},\ \bibinfo {pages} {246802} (\bibinfo {year}
  {2008})}\BibitemShut {NoStop}%
\bibitem [{\citenamefont {W\'ojs}\ \emph {et~al.}(2010)\citenamefont {W\'ojs},
  \citenamefont {M\"oller}, \citenamefont {Simon},\ and\ \citenamefont
  {Cooper}}]{Wojs-PhysRevLett.104.086801}%
  \BibitemOpen
  \bibfield  {author} {\bibinfo {author} {\bibfnamefont {A.}~\bibnamefont
  {W\'ojs}}, \bibinfo {author} {\bibfnamefont {G.}~\bibnamefont {M\"oller}},
  \bibinfo {author} {\bibfnamefont {S.~H.}\ \bibnamefont {Simon}}, \ and\
  \bibinfo {author} {\bibfnamefont {N.~R.}\ \bibnamefont {Cooper}},\ }\href
  {\doibase 10.1103/PhysRevLett.104.086801} {\bibfield  {journal} {\bibinfo
  {journal} {Phys. Rev. Lett.}\ }\textbf {\bibinfo {volume} {104}},\ \bibinfo
  {pages} {086801} (\bibinfo {year} {2010})}\BibitemShut {NoStop}%
\bibitem [{\citenamefont {Romers}\ \emph {et~al.}(2011)\citenamefont {Romers},
  \citenamefont {Huijse},\ and\ \citenamefont {Schoutens}}]{Romers-NJP-045013}%
  \BibitemOpen
  \bibfield  {author} {\bibinfo {author} {\bibfnamefont {J.}~\bibnamefont
  {Romers}}, \bibinfo {author} {\bibfnamefont {L.}~\bibnamefont {Huijse}}, \
  and\ \bibinfo {author} {\bibfnamefont {K.}~\bibnamefont {Schoutens}},\ }\href
  {http://stacks.iop.org/1367-2630/13/i=4/a=045013} {\bibfield  {journal}
  {\bibinfo  {journal} {New Journal of Physics}\ }\textbf {\bibinfo {volume}
  {13}},\ \bibinfo {pages} {045013} (\bibinfo {year} {2011})}\BibitemShut
  {NoStop}%
\bibitem [{\citenamefont {Moran}\ \emph {et~al.}(2012)\citenamefont {Moran},
  \citenamefont {Sterdyniak}, \citenamefont {Vidanovi\ifmmode~\acute{c}\else
  \'{c}\fi{}}, \citenamefont {Regnault},\ and\ \citenamefont
  {Milovanovi\ifmmode~\acute{c}\else \'{c}\fi{}}}]{Moran-PhysRevB.85.245307}%
  \BibitemOpen
  \bibfield  {author} {\bibinfo {author} {\bibfnamefont {N.}~\bibnamefont
  {Moran}}, \bibinfo {author} {\bibfnamefont {A.}~\bibnamefont {Sterdyniak}},
  \bibinfo {author} {\bibfnamefont {I.}~\bibnamefont
  {Vidanovi\ifmmode~\acute{c}\else \'{c}\fi{}}}, \bibinfo {author}
  {\bibfnamefont {N.}~\bibnamefont {Regnault}}, \ and\ \bibinfo {author}
  {\bibfnamefont {M.~V.}\ \bibnamefont {Milovanovi\ifmmode~\acute{c}\else
  \'{c}\fi{}}},\ }\href {\doibase 10.1103/PhysRevB.85.245307} {\bibfield
  {journal} {\bibinfo  {journal} {Phys. Rev. B}\ }\textbf {\bibinfo {volume}
  {85}},\ \bibinfo {pages} {245307} (\bibinfo {year} {2012})}\BibitemShut
  {NoStop}%
\bibitem [{\citenamefont {Thiebaut}\ \emph {et~al.}(2014)\citenamefont
  {Thiebaut}, \citenamefont {Goerbig},\ and\ \citenamefont
  {Regnault}}]{Thiebaut-PhysRevB.89.195421}%
  \BibitemOpen
  \bibfield  {author} {\bibinfo {author} {\bibfnamefont {N.}~\bibnamefont
  {Thiebaut}}, \bibinfo {author} {\bibfnamefont {M.~O.}\ \bibnamefont
  {Goerbig}}, \ and\ \bibinfo {author} {\bibfnamefont {N.}~\bibnamefont
  {Regnault}},\ }\href {\doibase 10.1103/PhysRevB.89.195421} {\bibfield
  {journal} {\bibinfo  {journal} {Phys. Rev. B}\ }\textbf {\bibinfo {volume}
  {89}},\ \bibinfo {pages} {195421} (\bibinfo {year} {2014})}\BibitemShut
  {NoStop}%
\end{thebibliography}%

\end{document}